\documentclass[12pt]{article}
\pdfoutput=1
\usepackage{amssymb, amsmath,amsfonts}
\usepackage{multirow}
\usepackage{mathrsfs}
\usepackage{array}
\usepackage{cite}
\usepackage{booktabs}
\usepackage{tikz}
\usepackage{pgfplots}
\usepackage{float}
\usepackage{subfigure, graphicx}
\usepackage[pdftex, bookmarks=true,colorlinks,linkcolor=red,urlcolor=blue,citecolor=blue]{hyperref}
\usepackage{cite}
\usepackage{slashed}
\usepackage{tabularx}

\usepackage{tikz}

\makeatletter\@addtoreset{equation}{section}\makeatother

\def\be{\begin{equation}}
\def\ee{\end{equation}}
\def\bea{\begin{eqnarray}}
\def\eea{\end{eqnarray}}

\newcommand{\nn}{\nonumber}

\def\Dslash{\,\,{\raise.15ex\hbox{/}\mkern-12mu D}}
\def\Dbarslash{\,\,{\raise.15ex\hbox{/}\mkern-12mu {\bar D}}}
\def\delslash{\,\,{\raise.15ex\hbox{/}\mkern-9mu \partial}}
\def\delbarslash{\,\,{\raise.15ex\hbox{/}\mkern-9mu {\bar\partial}}}
\def\pslash{\,\,{\raise.15ex\hbox{/}\mkern-9mu p}}
\def\calDslash{\,\,{\raise.15ex\hbox{/}\mkern-12mu {\cal D}}}

\makeatletter\@addtoreset{equation}{section}\makeatother

\renewcommand{\title}[1]{\vbox{\center\LARGE{#1}}\vspace{5mm}}
\renewcommand{\author}[1]{\vbox{\center#1}\vspace{5mm}}
\newcommand{\address}[1]{\vbox{\center\em#1}}

\def\arXiv#1{\href{http://arxiv.org/abs/#1}{arXiv:#1}}
\def\arXiv#1#2{\href{http://arxiv.org/abs/#1}{arXiv:#1}}

\def\doi#1#2{\href{http://doi.org/#1}{#2}}

\begin{document}

\unitlength = .8mm

\begin{titlepage}
\vspace{.5cm}
 
\begin{center}
\hfill \\
\hfill \\
\vskip 1cm

\title{Properties of gapped systems in AdS/BCFT
}
\vskip 0.5cm
{Yan Liu$^{\,a,b}$}\footnote{Email: {\tt yanliu@buaa.edu.cn}}, 
{Hong-Da Lyu$^{\,a,b}$}\footnote{Email: {\tt hongdalyu@buaa.edu.cn}}  and {Jun-Kun Zhao$^{\,c,a}$}\footnote{Email: {\tt junkunzhao@itp.ac.cn}}

\address{${}^{a}$Center for Gravitational Physics, Department of Space Science\\ and International Research Institute
of Multidisciplinary Science, \\Beihang University, Beijing 100191, China}

\address{${}^{b}$Peng Huanwu Collaborative Center for Research and Education, \\Beihang University, Beijing 100191, China}

\address{${}^{c}$CAS Key Laboratory of Theoretical Physics, Institute of Theoretical Physics,\\ Chinese Academy of Sciences, Beijing 100190, China}

\end{center}
\vskip 1.5cm

\abstract{We study the conductivities and entanglement structures of two different holographic gapped systems at zero density in the presence of  boundaries within AdS/BCFT. The first gapped system is described by the Einstein-scalar gravity and the second one is the dual of AdS soliton geometry. We show that in both these two systems the bulk and boundary conductivities along the spatial  direction of the boundary of BCFT are trivial. For the first system, when we increase the size of the subsystem the renormalized entanglement entropy is always non-negative and monotonically decreasing  with discontinuous, or continuous, or smooth behavior, depending on the effective tension of the brane. While for the AdS soliton with a boundary, the renormalized entanglement entropy only exhibits a discontinuous drop when we increase the size of the subsystem. 
}
\vfill

\end{titlepage}

\begingroup 
\hypersetup{linkcolor=black}
\tableofcontents
\endgroup





\section{Introduction}

The AdS/CFT correspondence, also known as holographic duality, provides a novel way to study strongly correlated quantum  systems in terms of weakly coupled gravity. In particular, it can describe strongly correlated gapped systems in terms of gravity duals. One class of such models at zero density include the Girardello-Petrini-Porrati-Zaffaroni gapped geometry \cite{Girardello:1999hj}, AdS soliton \cite{Witten:1998zw}, AdS with cutoffs in IR \cite{Erlich:2005qh} and so on. Another class of models are for finite density systems with translational symmetry-breaking effects; see, e.g., \cite{Kiritsis:2015oxa}.  

Physical systems in the real world often have boundaries, and the boundary effects play important roles, ranging from D-branes in string theory to topological states in condensed matter physics. One well-known example of the topological states in a condensed matter system is the topological insulator, which is gapped in the bulk while nontrivial gapless charged excitations exist on the boundary \cite{ti-kane}. 
Constructing a holographic model of topological insulators 
is a difficult question for the bottom-up holography.\footnote{Previous attempts to study the holographic model of topological insulator from the top-down approach in the probe approximation (via probe branes) include, e.g., \cite{Hoyos-Badajoz:2010etp, Rozali:2012gf}.}  
Because of the difficulty and as a preliminary step with the hope that we could 
obtain some important hints toward constructing a model of topological insulators, we start from
a simpler while nontrivial question to analyze what happens to a holographic gapped system in the presence of a boundary. In condensed matter physics, both the bulk of the topological insulator and normal
insulator have a hard gap in the band structure, while they have different boundary states.
The holographic gapped system without a boundary shows a gap. Though there has been
no evidence showing that it is topologically nontrivial, as the Hilbert space has fundamentally
changed compared to the weakly coupled field theory, we still need to check the boundary states
to see if it is topologically trivial or nontrivial. Therefore, the question on the properties 
 for such a system in the presence of a  boundary is natural and important.

We study this problem in the framework of AdS/BCFT. In AdS/CMT there are few studies on the effects of ``soft" boundaries by considering matter fields with spatially dependent profiles which separate two different phases; see, e.g., \cite{Horowitz:2011dz, Ammon:2016mwa}.  Here, we use AdS/BCFT to describe a ``hard" boundary of the holographic system which might make the model be more realistic. AdS/BCFT allows us to study the properties of field theories with boundaries from the holographic dual. 
In AdS/BCFT, the bulk geometry terminates at the end-of-the-world (EOW) brane such that the boundary of the EOW brane near AdS coincides with the boundary of BCFT \cite{Takayanagi:2011zk, Fujita:2011, Karch:2000gx}. AdS/BCFT has been actively explored during the past decade. A far from complete list includes applications to condensed matter physics \cite{Fujita:2012, Melnikov:2012tb}, cosmology \cite{Antonini:2019qkt}, black hole physics \cite{Geng:2021mic, Suzuki:2022xwv, Yadav:2022mnv},  quantum information \cite{Seminara:2018pmr}, and so on. However, so far, the studies of AdS/BCFT have been mainly limited to critical gapless systems with boundaries. Studies of gapped systems in such a framework might provide more insights on the properties of strongly interacting quantum field theories with boundaries.  

The purpose of this paper is to study the properties of gapped systems in the presence of boundaries in the framework of AdS/BCFT. We will focus on the vacuum states of the first class of models as mentioned in the first paragraph at zero temperature and zero density. Here, we consider two different holographic models of gapped systems. The first one is  the gapped geometry in Einstein-scalar theory. We choose the Neumann  boundary condition for the fields on the EOW brane. Taking a proper scalar potential term localized on the brane, we can get a consistent background for the gapped geometry with an EOW brane. Then we will study the transport properties and entanglement entropies of the BCFT.  The second gapped system is described by the AdS soliton  \cite{Witten:1998zw}. The AdS soliton can be obtained by analytic continuation of the AdS Schwarzschild black hole. At finite temperature, there is a first-order phase transition between the AdS Schwarzschild black hole and the AdS soliton, which describes the confinement-deconfinement phase transition. There is a compact spatial dimension in the AdS soliton which sets the scale of the transition. 
We consider the presence of a boundary for the dual field theory of the AdS soliton 
along one noncompact spatial direction and study its transport properties and entanglement entropies. We will make comparisons on the profiles and the properties between these two different gapped systems in the presence of boundaries.   

Our paper is organized as follows. In section \ref{sec2}, we first construct a gapped system in the presence of a boundary in Einstein-scalar theory using AdS/BCFT, and then study its conductivity along the spatial direction of the boundary as well as its entanglement entropy. In section \ref{sec3}, we study the properties of a gapped system which is described by the AdS soliton in AdS/BCFT. We summarize our results in section \ref{sec4} and discuss the possible open questions. Some calculation details are collected in the appendices.

\section{A gapped system in AdS$_4$/BCFT$_3$}
\label{sec2}

In this section, we study the holographic gapped system with boundaries in the Einstein-scalar gravity and consider its properties in the  framework of AdS/BCFT \cite{Takayanagi:2011zk, Fujita:2011}. We focus on the case of three-dimensional field theories with two-dimensional boundaries and it is straightforward to generalized to other dimensions.

\begin{figure}[h]
\begin{center}
\begin{tikzpicture}[scale=0.7]
\draw [fill=blue!10] (0,0)--(7,0)--(10,3)--(3,3);
\draw [fill=green!5]  (0,0)
to[out=150, in=270](-3,5)--(0,8)
to[out=-90, in=-210](3,3);

\draw [fill=red] (0,0)--(3,3) node[anchor=east,midway]{ $P$ };

\draw[black] (5.0, 1.5)  node{ $M$ };
\draw[black] (-1.0, 4.0)  node{ $Q$ };
\draw[black] (5.0, 6.0)  node{ $N$ };
\draw[black] (0, 0);

\draw[-latex, very thick, blue, opacity=0.7] (7, 0)--(8.5, 1.5)   node[anchor=west, near end]{$ ~\{ t,\,y\} $ };
\draw[-latex, very thick, blue, opacity=0.8] (0, 0)--(8.5, 0) node[anchor=north, at end]{ $x$ };
\draw[-latex, very thick, opacity=0.8 ] (11, 2)--(11, 5) node[anchor=west, midway]{ $u$ };
\end{tikzpicture}
\end{center}
\vspace{-0.3cm}
\caption{\small The configuration under consideration. The field theory lives in the manifold $M$ with boundary $P$. The dual gravity lives in the bulk $N$ with boundary $Q$. }
\label{fig:cf}
\end{figure}
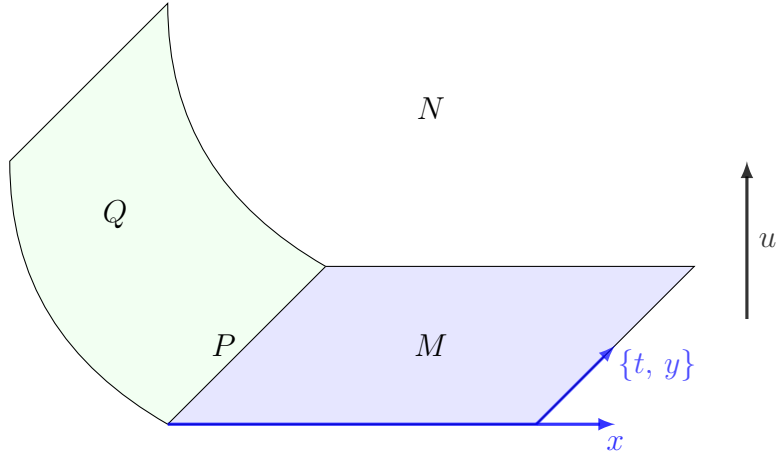

The configuration under consideration is shown in Fig. \ref{fig:cf}. The three-dimensional boundary field theory is defined on the manifold $M$ with boundary $P$ along the $y$ direction. The gravity dual lives in the bulk $N$ with the EOW brane $Q$ which anchors to the BCFT boundary $P$. Note that $u$ is the holographic direction and the boundary $M$ lives at $u=0$. 

We consider the Einstein-scalar gravitational theory
\be
S_\text{bulk}=S_N +S_Q  \,,
\ee
with
\bea
\label{eq:action1}
\begin{split}
S_N&=\int_N d^4x\sqrt{-g}\,\bigg[\frac{1}{2\kappa^2}\bigg(R+6-\frac{1}{2}(\partial\phi)^2
    -V(\phi)\bigg)-\frac{Z(\phi)}{4e^2}F^2
    \bigg]  \,,\\
S_{Q}&=\int_Q d^3x\sqrt{-\gamma}\, \bigg[\frac{1}{\kappa^2}(K-T\big)+v(\phi)
\bigg] \,,\\
\end{split}
\eea
where the bulk gauge field $A_a$ is dual to the electric current on the boundary and it has field strength $F_{ab}=\partial_a A_b-\partial_b A_a$. $\kappa$ and $e$ are the gravitational constant and the bulk gauge coupling constant, respectively. Note that the scalar field $\phi$ is real. The induced metric on the EOW brane $Q$ is denoted as $\gamma_{\mu\nu}$, where $K$ and $T$ are the extrinsic curvature and the tension of the EOW brane $Q$, respectively. Note that on $Q$ we also consider a potential term $v(\phi)$, and it contributes to the effective tension of the brane.

 We set $2\kappa^2=e^2=1$. 
The equations of motion  in the bulk $N$ are
\bea
\begin{split}
R_{ab}-\frac{1}{2}g_{ab}\big(R+6\big)-\frac{1}{2}T_{ab}&=0 \,,\\
\nabla_b\big(Z(\phi) F^{ba}\big)
&=0\,,\\
\nabla_{a}\nabla^{a}\phi-\frac{\partial_\phi Z(\phi)}{4}F^2-\partial_\phi V(\phi)&=0 \,, 
\end{split}
\eea
where
\bea
T_{ab}&=&Z(\phi)\bigg[F_{ac}F_{b}^{~c}-\frac{1}{4}g_{ab}F^2\bigg]
         +\nabla_{a}\phi\nabla_{b}\phi-g_{ab}\bigg[\frac{1}{2}(\partial\phi)^2+V(\phi)\bigg] \,. \nn
\eea

The equations of motion on the EOW brane $Q$ can be obtained from the variations. The variation for metric fields, scalar field, and vector field yields
\bea
\begin{split}
\delta S\Big{|}_Q&=
  \int_Qd^3x\sqrt{-\gamma}\,\frac{1}{2\kappa^2}\left[K_{\mu\nu}-(K-T)\gamma_{\mu\nu} -\frac{1}{2} v(\phi) \gamma_{\mu\nu}
  \right] \delta\gamma^{\mu\nu}
 \\
&+\int_Q d^3x\sqrt{-\gamma}\,\left[-n^a\nabla_a\phi+v'(\phi) 
\right]\,\delta\phi \\
&+\int_Q d^3x \sqrt{-\gamma}\,n_a\left( -\frac{Z(\phi)}{e^2}F^{ab}\right)\,\delta A_b\,.
\end{split}
\eea
Note that $n^a$ is the outward unit vector of $Q$. Here, $\gamma_{\mu\nu}$ should be understood as the metric from the Gaussian normal coordinate on the EOW brane. Following the standard AdS/BCFT, we impose Neumann boundary condition on $Q$.\footnote{AdS/BCFT with a Dirichlet boundary condition or mixed boundary condition can be found in, e.g.,  \cite{Chu:2017aab, Miao:2018qkc, Guijosa:2022jdo}. It would be interesting to consider gapped systems with generalized boundary conditions in AdS/BCFT.} 
Then we obtain the following equations on $Q$: 
\bea{\label{eq:Qbc}}
\begin{split}
K_{\mu\nu}-(K-T)\gamma_{\mu\nu}-\frac{1}{2}v(\phi)\gamma_{\mu\nu}&=0\,, \\
n^{a}\partial_a\phi-\partial_\phi v(\phi)&=0\,,\\
n_aF^{ab}&=0\,.
\end{split}
\eea
Note that we choose the Dirichlet boundary condition on $M$ and $P$.

\subsection{Zero temperature ground state}
\label{subsec:groundstate}

We focus on the vacuum solution at zero temperature and zero density and consider the following ansatz of the metric fields, scalar, and vector fields: 
\bea\label{eq:ansatz d4}
ds^2=\frac{1}{u^2}\bigg[-dt^2+dx^2+dy^2+\frac{du^2}{f(u)}\bigg] \,,~~~ \phi=\phi(u)\,,~~~A_a=0\,.
\eea
Near the AdS boundary, i.e., $u\to 0$, the metric field $f(u)\to 1$. The IR regime is $u\to\infty$.  

The equations of motion for the system in $N$ are
\bea
\begin{split} 
\frac{V-6}{u^2 f}+\frac{6}{u^2}-\frac{1}{2} \phi'^2&=0\,, \\
\phi'^2-\frac{2f'}{uf}&=0\,,\\
\phi''+\phi'\left(
\frac{f'}{2f}-\frac{2}{u}\right)-\frac{\partial_\phi V}{u^2 f}&=0\,.
\end{split}
\eea

We have the bulk solution which satisfies the gapped spectrum condition 
\be
\label{eq:bgsol1}
f(u)=1+a_0 u^n\,,~~~\phi(u)=\frac{2\sqrt{2}}{\sqrt{n}}\,\text{arcsinh}\big[\sqrt{a_0}u^{\frac{n}{2}}\big]
\ee
with 
\be\label{eq:bgsp}
V(\phi)=(n-6)\,\Big(\sinh\Big[
\frac{\sqrt{n}}{2\sqrt{2}} \phi \Big]\Big)^2\,.
\ee 
Note that $a_0>0$ and can be set to be $1$ using the scaling symmetry $u\to \lambda u,\, (t,x,y)\to \lambda(t,x,y),\, (f,\phi)\to (f,\phi)$. In the following, we set $a_0=1$. 

In the IR region, i.e., $u\to\infty$, from the solution \eqref{eq:bgsol1} we have $f(u)\to u^n$. It is known that this kind of geometry has a gapped spectrum for probe fields when $n\geq 2$ \cite{Liu:2013una}.
Additionally, 
in the deep IR, i.e.,  $u\to\infty$, the Ricci scalar is divergent (except the $n=4$ case, where the Ricci scalar is finite while the Kretschmann scalar is divergent),  
from which we know that there is a curvature singularity for the solution \eqref{eq:bgsol1}. Nonetheless, the singularity is physically acceptable if the Gubser criterion is satisfied \cite{Gubser:2000nd, Charmousis:2010zz}, which constrains 
$n \leq 6$.\footnote{Note that the strong energy condition requires $n \leq 6$, while the null energy condition does not put any constraint on the system.} In the following, we will focus on the cases $n\in [2, 6].$

Near the AdS boundary, we have $\phi\to0$ and  $V(\phi)=\frac{n(n-6)}{8}\phi^2+\cdots$. This gives the effective mass of scalar field $m^2=\frac{n(n-6)}{4}$, which is always above the BF bound for arbitrary $n$. Here, we focus on the parameter regimes $2\leq n\leq 6$. The scalar field near the AdS boundary behaves as  
\be \phi\to  \frac{ 2\sqrt{2} }{\sqrt{n}} u^{n/2}\,\Big(1-\frac{1}{6} u^n +\frac{3}{40} u^{2n}+\cdots\Big)\,.
\ee For the parameters we are interested in, i.e., $n\in \,[2, 6]$, $\phi$ is dual to operators of dimension $n/2$. The dual system is a $Z_2$ spontaneously symmetry broken state.\footnote{Note that, for $n\in \,[2, 5]$, both quantizations are possible, $\phi$ could also be viewed as being dual to an operator with conformal dimension $(6-n)/2$, and this seems to be an unphysical case since the dual theory has a deformation with a scalar source which does not produce any response.}

For simplicity, we suppose the manifold $M$ is restricted to be a half infinite plane with coordinates $t,y$, and $x$ with $x\geq 0$. Assuming the boundary $Q$ is parametrized as $x(u)$, the spacelike unit vector $n^a$ normal to the boundary $Q$ (outward direction) is given by
\bea
\label{eq:normal1}
(n^t,n^x,n^y,n^u)=\bigg(0,~~\frac{-u}{\sqrt{1+f(u)x'(u)^2}},~~0,~~\frac{u f(u)x'(u)}{\sqrt{1+f(u)x'(u)^2}}\bigg)\,.
\eea
The extrinsic curvature $K_{ab}$ can be obtained from $K_{ab}=h_a^{~c}h_b^{~d}\nabla_c n_d$ where $h_{ab}=g_{ab}-n_a n_b$. 
Note that, since the coordinates here are not the Gaussian normal coordinate of the EOW brane, we should use $h_{ab}$ to calculate the boundary equations. In the end, the boundary conditions (\ref{eq:Qbc}) result in the following constraints on the EOW brane $Q$:  
\bea
\label{eq:bc}
\begin{split}
x''+\frac{f'}{2f}x' &=0   \,, \\
x'+\frac{\big(2T-v(\phi) \big)\sqrt{1+fx'^2}}{4f} &=0  \,, \\
n^{u}\partial_u\phi-\partial_\phi v(\phi) &=0\,.
\end{split}
\eea

From the above equations, we find the  solution for the EOW brane $Q$:
\bea
\label{eq:bgeol2}
x&=&c\, u\, {}_2F_1\left[\frac{1}{2},\frac{1}{n}, 1+\frac{1}{n}, - u^n \right]\,,\\
\label{eq:scalarpot2}
v(\phi)&=&2T+\frac{4c}{\sqrt{1+c^2}}\cosh
\left[\frac{\sqrt{n}}{2\sqrt{2}}\phi 
\right]\,,
\eea 
where $c$ is a real integration constant. The first equation parametrizes the profiles of the EOW brane, while the second equation is a potential term for the scalar field on $Q$ which should be thought of as an input quantity to  determine the profiles of the system.  

\begin{figure}[h]
\begin{center}
\includegraphics[width=0.62\textwidth]{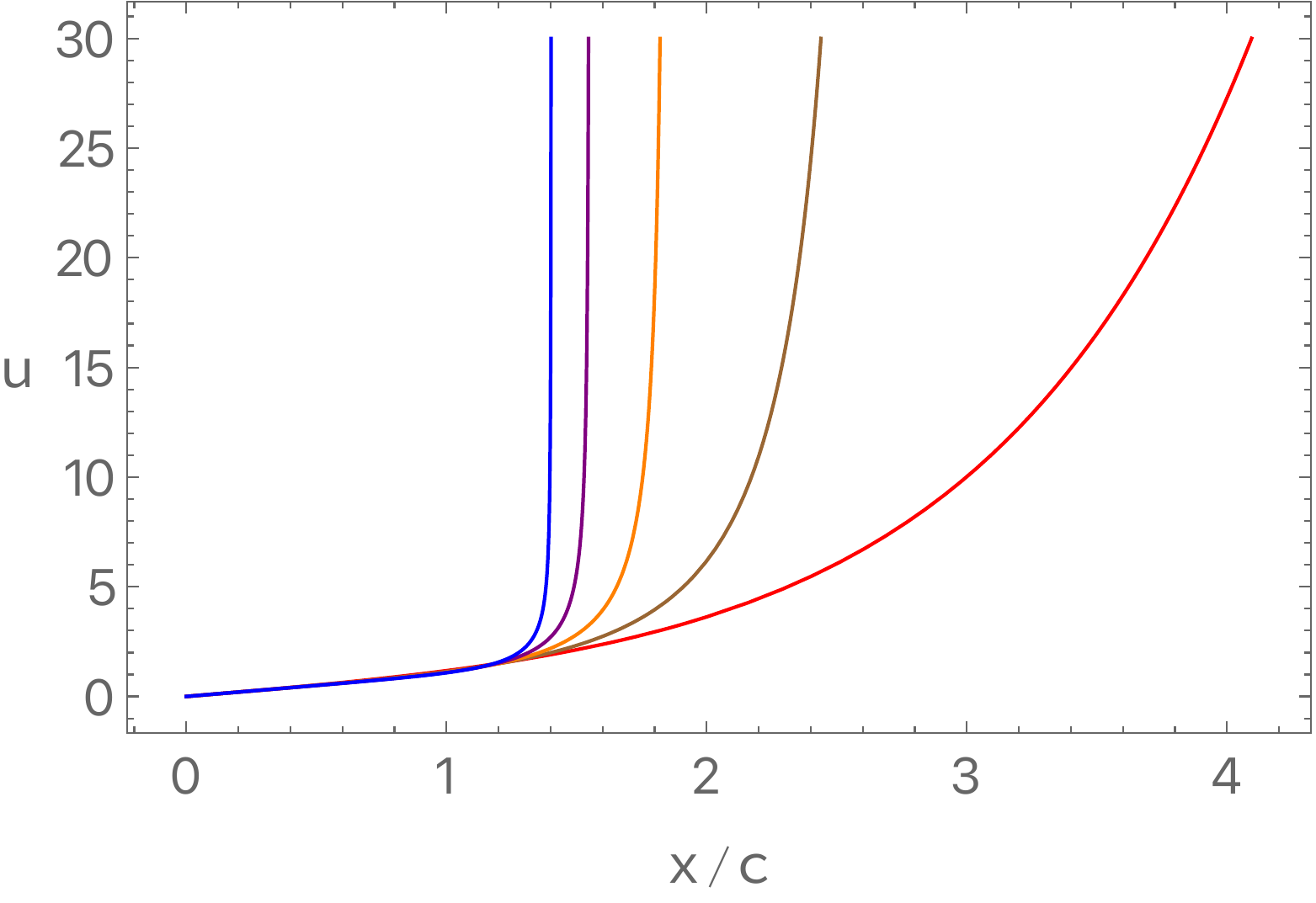}
\end{center}
\vspace{-0.3cm}
\caption{\small The plot for the profile of the EOW brane $Q$ as a function of  $x/c$ when $c\neq 0$ for $n=2$ (red), $\,3$ (brown), $\,4$ (orange), $\,5$ (purple), $\,$and $6$ (blue). When $c<0$, the EOW brane $Q$ extends along negative $x$, while it extends along positive $x$ when $c>0$. }
\label{fig:config}
\end{figure}

Equations (\ref{eq:bgsol1}) and (\ref{eq:bgeol2}) are the background solutions of the gravitational system.
The profile of the EOW brane $Q$, which is described by $x(u)$ in \eqref{eq:bgeol2}, is independent of the parameter $T$  while it depends on the effective tension $T-v(\phi)/2$, i.e.,  the contribution of the potential of the scalar field that is parametrized by the parameter $c$. 
When $c=0$, the profile of the EOW brane $Q$ is trivial, and it is given by $x=0$. When $c\neq 0$,  different from the case without a scalar field that was first studied in \cite{Takayanagi:2011zk}, the profile of $Q$ here is nonlinear in $u$. Near the AdS boundary, i.e., $u\to 0$, we have linear behavior at leading order:  
\be 
\frac{x}{c}= u- \frac{1}{2(n+1)} u^{n+1} + \frac{3}{8(2n+1)}u^{2n+1}+\cdots ,
\ee
while in  the deep IR, i.e., $u\to\infty$, we have
\be 
\frac{x}{c} =
\begin{cases}
\log(2u)+\frac{1}{4 u^2} \cdots   &\quad\quad  \textrm{if $n=2$\,,}  \\[2ex]
\frac{1}{\sqrt{\pi}}\Gamma \left(\frac{1}{2}-\frac{1}{n} \right) \Gamma \left( 1+\frac{1}{n}\right)- \frac{2 n}{n-2} \frac{ \Gamma (1+\frac{1}{n} )}{ \Gamma  (\frac{1}{n}) } u^{1-\frac{n}{2}}+\cdots    &\quad\quad  \textrm{if $n>2$\,,}
\end{cases}
\ee
These expressions indicate that, near the boundary $P$ of BCFT, the profile of the EOW brane $Q$ is linear in $x$ with a  slope $1/c$. When $u\to \infty$, the EOW brane approaches infinity for $n=2$, depending on the sign of $c$, while it approaches a constant $x_m$ for $n=3,4,5,6$. 
Fig. \ref{fig:config} shows the profiles of the EOW brane as a function of $x/c$ at different values of $n$. The particular properties of the profiles for the EOW brane will play important roles in the calculations of the entanglement entropy that we study in section \ref{subsec:2ee}. 

With the profile (\ref{eq:bgeol2}) of $Q$, the normal vector \eqref{eq:normal1} on $Q$ can be simplified as 
\bea
n^a=\bigg(0,~~\frac{-u}{\sqrt{1+c^2} },~~0,~~\frac{c\, u \sqrt{1+
u^n } }{ \sqrt{1+c^2} } \bigg)\,.
\eea
Then we can obtain the projection tensor 
\be
h_{ab}dx^adx^b=\frac{1}{u^2}\left[-dt^2+dy^2+\frac{1}{1+c^2}\bigg(c\, dx +\frac{du}{ \sqrt{f}}\bigg)^2\right]\,.
\ee
One can check that the trace of extrinsic curvature on the EOW brane,  $K=\frac{-3c }{\sqrt{1+c^2} }\,  \sqrt{1+u^n}$, is divergent near the singularity of geometry $N$ when $c\neq 0$. From the induced metric 
\be
\label{eq:im}
\tilde{\gamma}_{\mu\nu}dx^\mu dx^\nu\Big{|}_Q=\frac{1}{u^2}\left[-dt^2+dy^2+\frac{1+c^2}{f}d u^2\right]\ee 
in the coordinates $\{t, y, u\}$, 
we know that, on the EOW brane, the metric is asymptotic to AdS$_3$ in UV with AdS radius $\sqrt{1+c^2}$, which is different from the one in $N$. Similar to the gapless system in AdS/BCFT, which is a pure AdS on the EOW brane \cite{Takayanagi:2011zk}, the induced metric on $Q$ is also asymptotic AdS. The intrinsic curvature from the induced metric on the EOW brane $Q$ is, in general, divergent except $n=3$, where the  Kretschmann scalar is divergent. Nonetheless, it is physically acceptable following the arguments in \cite{Gubser:2000nd, Charmousis:2010zz}, since it is believed that all the singularities are resolvable after considering extra degrees of freedom that do not affect any calculations here. 

One particular interesting case is $n=2$. In this case, all the above formulas can be simplified, and we just collect them here for later use: 
\bea\label{eq:ir2}
\begin{split}
f(u)& =1+ 
u^2  \,, ~~~~
\phi(u)=2\,\text{arcsinh}\big[
u \big]\,,~~~~x(u)=c\, 
\text{arcsinh}\big[ 
u \big]
\,,\\
 V(\phi)&=-4\,\left(\text{sinh}\left[\frac{\phi}{2} \right]\right)^2\,
, ~~~~
v(\phi)=2T+\frac{4c}{\sqrt{1+c^2}}\text{cosh}\left[\frac{\phi}{2}\right]\,.
\end{split}
\eea

\subsection{Conductivity}
\label{subsec:con}

In the previous subsection, we have constructed the gapped geometry in the presence of a boundary and found the solutions (\ref{eq:bgsol1}) and (\ref{eq:bgeol2}) with proper choices of the potential terms for the scalar fields in the bulk $N$ \eqref{eq:bgsp} and on the EOW brane $Q$ \eqref{eq:scalarpot2}. In this subsection, we will study its conductivity along the spatial direction $y$ of the boundary $P$. 

We consider the linear fluctuations of the gauge fields 
\be
\label{eq:gauflu}
\delta A_i(t, x, u)=\int\frac{d\omega}{2\pi} a_i(\omega, x, u) e^{-i\omega t}\,.
\ee
We are interested in the conductivity along the $y$ direction. It turns out that the equation of motion for $a_y$ decouples from other fields. The fluctuation equation for $a_y$ in $N$ is 
\bea
\label{eq:fulay1}
a_y''+\left(\frac{f'}{2f}+\frac{\phi' \partial_\phi Z}{Z}\right)a_y'+\frac{\omega^2
+\partial_x^2}{f} a_y&=0\,, 
\eea
and the equation for $a_y$ on the boundary $Q$ is 
\be\label{eq:be}
(-\partial_x a_y +f x' \partial_u a_y)\Big{|}_Q=0\,.
\ee
Here, the prime denotes the derivative with respect to the radial coordinate $u$. Now we have a boundary value problem for the partial differential equation \eqref{eq:fulay1}. 

The solution of the above equations depend on whether $c$ equals zero or not, and we first focus on the case with nonzero $c$. 
When $c\neq 0$, we can solve \eqref{eq:fulay1} by using the separation of variables. We choose 
\be
\frac{Z'}{Z}=\frac{\alpha}{\sqrt f}\,,
\ee 
where the prime is the derivative with respect to $u$, i.e., 
\be
\label{eq:conZ}
\setlength\arraycolsep{1pt}
Z=\exp{\left(\alpha\Big(\sinh\left[\frac{\sqrt{n}}{2\sqrt 2}\phi\right]\Big)^{\frac{2}{n}} \,  {}_2F_1\left[\frac{1}{2},\frac{1}{n},1+\frac{1}{n},-\Big(\sinh\left[\frac{\sqrt{n}}{2\sqrt 2}\phi\right]\Big)^2\right]\right)}\,.
\ee
Note that we have normalized $Z\to 1$ near the AdS boundary. When $n=2$, the above result can be further simplified as $Z= e^{\alpha \phi/2}$. For other values of $n$, we have $u\to \infty$, $Z\sim e^{\alpha x_m}$; i.e., $Z$ approaches a constant value in the deep IR. 

With this choice of $Z$, we find the solution of \eqref{eq:fulay1} with boundary equation \eqref{eq:be} is 
\be 
\label{eq:solay}
a_y=e^{b x +\frac{b}{c^2} x(u)-i\omega t}\,.
\ee
The second term $x(u)$ in the exponential should be viewed as the solution in \eqref{eq:bgeol2}, and, in this way, $a_y$ is a function explicitly depending on the variables $x, u,$ and $t$. In \eqref{eq:solay}, we have (when $|\omega| < \frac{|\alpha|}{2\sqrt{1+c^2}}$) 
\be
b=\frac{c}{2(1+c^2)}
\left(-\alpha\pm\sqrt{\alpha^2-4(1+c^2)\omega^2}\right)\,.
\ee  
For $|\omega| <\frac{|\alpha |}{2\sqrt{1+c^2}}$, the solution (\ref{eq:solay}) is real and normalizable\footnote{This is true only for the cases $n>2$ and $c\alpha>0$, and the following discussion should apply for these cases. When $n=2$, the field $a_y$ is divergent at either $x\to \infty$ or $x\to -\infty$, and we do not have a reliable solution yet. One might  expect that the conclusions below are also true for the case of $n=2$. } for both choices of $b$. However, for $\omega>\frac{|\alpha|}{2\sqrt{1+c^2}}$,   the sector with
\be
b=\frac{c}{2(1+c^2)}\left(-\alpha + i\sqrt{4(1+c^2)\omega^2-\alpha^2}\right) 
\ee
describes the infalling wave. 
 Following \cite{Kiritsis:2015oxa, Yan2018}, we use the analytic  continuation from $|\omega| > \frac{|\alpha |}{2\sqrt{1+c^2}}$ to $|\omega| < \frac{|\alpha |}{2\sqrt{1+c^2}}$ to fix 
 \be
 \label{eq:b}
 b=\frac{c}{2(1+c^2)}\left(-\alpha-
 \sqrt{\alpha^2-4(1+c^2)\omega^2}\right)\,.
 \ee
 
From the above solution, i.e. \eqref{eq:solay} with \eqref{eq:b}, we can compute the conductivity. When $u\to 0$, we have 
\be a_y=e^{bx-i\omega t}\,\left(1+ \frac{b}{c}u+\mathcal{O}(u^2)\right)\,. \ee

Note that, on $M$, we choose the Dirichlet boundary condition for the gauge field (i.e., fixing the source) and we do not need to include any counterterm for the gauge field.  
We have the on-shell action for the gauge field:
\be
S_M=-\int dt dx dy\,\sqrt{-\gamma}\, Z A_\nu F^{u\nu}n_u
\ee
where $\gamma$ is the induced metric on $M$ while $n_u$ is an outward-pointing unit vector of $M$, i.e., $\gamma_{\mu\nu}=\text{diag} \big(-1/u^2, 1/u^2, 1/u^2\big)$ and $n^u=-u \sqrt{f}$. 
In the case with the fluctuations of the gauge field along the $y$ direction, we have  
\be
S_M=\int dtdxdy\, Z a_y \partial_u a_y=
\int dtdxdy\,  a_y^{(0)}a_y^{(1)}\,
\ee
from which we have the retarded Green's function on $M$:
\be
G_R=\frac{a_y^{(1)}}{a_y^{(0)}}\,.
\ee

Therefore, we have conductivity in $M$: 
\be
\sigma_{y}= \frac{1}{i\omega}\frac{a_y^{(1)}}{a_y^{(0)}}=\frac{b}{i\omega c}\,.
\ee
For $\omega<\frac{|\alpha |}{2\sqrt{1+c^2}}$, $b$ is real; this means that $\sigma_y$ is pure imaginary. 
The DC conductivity in $M$ along the $y$ direction can be obtained from the real part:
\be\sigma_\text{DC}=\lim_{\omega\to 0} \text{Re} [\sigma_y] =0\,.\ee
Note that we have assumed $\alpha<0$ for simplicity and used the fact that $b\simeq -\frac{c}{\alpha}\omega^2$ when $\omega\to 0$, which means that there is no pole for $\sigma_y$ at $\omega\to 0$.\footnote{Note that, when $\alpha>0$, from \eqref{eq:b} one concludes that there is a pole at $\omega=0$. However, from the experimental point of view, one needs  to consider the subtle commutability between the two limits $T\to 0$ and $\omega\to 0$. Nevertheless, now we have $\lim_{\omega\to \epsilon^+}\text{Re}[\sigma_y]=0 $, and one might naively take it as an insulator for any $\alpha$.}
The gap of the conductivity is given by $\frac{|\alpha |}{2\sqrt{1+c^2}}$.\footnote{One might need to calculate the conductivity along the $x$ direction to confirm if it is also gapped for $\sigma_x$, and we will not discuss this here.}  
When $\omega >\frac{|\alpha |}{2\sqrt{1+c^2}}$, we have a nonzero conductivity with 
$\text{Re}[\sigma_y]=\frac{1}{2\omega (1+c^2)}\sqrt{4(1+c^2)\omega^2-\alpha^2}$.  

For the boundary $P$, we have not considered any dynamics of the gauge field on the EOW brane $Q$. The gauge field on $Q$ should be understood as the induced gauge field of $A_a$ in the bulk. Since the induced metric on $Q$ is asymptotic AdS$_3$, it is known from AdS$_3$/CFT$_2$ \cite{Jensen:2010em} that the expansion of the gauge field near $P$ depends on the action in the bulk. For a gauge field with  a canonical kinetic term\footnote{In the presence of a Chern-Simons term, the expansion will be slightly different and depends on the level \cite{Jensen:2010em, Andrade:2011sx}. However, the dual current is no longer conserved, and we will not consider this case here.}
we have \cite{Jensen:2010em, Faulkner:2012gt}
\be\label{eq:exax2}
a_y \sim a_y^{(r)}\log(u)+a_y^{(s)}+\cdots\,,
\ee
where $a_y^{(r)}$ is the response of the dual operator while $a_y^{(s)}$ can be understood as the source term. Note that there is a scaling anomaly and the definition of the source depends on the Landau pole of the theory \cite{Faulkner:2012gt}.  
  Along $P$,  we evaluate the solution (\ref{eq:solay}) on $Q$ and obtain  
\be
a_y\Big{|}_Q=e^{-i\omega t}\left(1+b(c+\frac{1}{c})u+\cdots\right)\ee when $u\to 0.$ Comparing to \eqref{eq:exax2}, we know that the Green's function is completely trivial, and, therefore, we have $\sigma=0$ on $P$.


When $c=0$, the boundary equation \eqref{eq:be} can be further simplified as 
$\partial_x a_y=0$ on $Q$. 
In this case we have solution $a_y=a_y(u)$ which is solution \eqref{eq:fulay1} with $\partial_x^2 a_y=0$. In appendix \ref{app:sch}, we analyze the solution of this equation by writing it into a Schr\"{o}dinger problem and show that it indeed has a gapped spectrum. Repeating the previous study along $P$, one concludes that the conductivities are trivial in both $M$ and $P$.

Our study shows that, for the holographic insulator in the presence of a boundary, the conductivity on the boundary is also trivial. This indicates that strong correlation would not make a trivial insulator topologically nontrivial.\footnote{In the literature of condensed matter physics, there are also examples of a topological insulator with gapped boundary states \cite{Witten:2015aba, Seiberg:2016rsg}, and it would be interesting to be attempt to make contact with these field theories.} To obtain a topological insulator, it seems that one has to add more dynamics of the gauge field on $Q$, and we leave this possibility for future investigation.


\subsection{Entanglement entropy}
\label{subsec:2ee}

 Entanglement entropy is an important physical quantity in quantum many-body systems \cite{Nishioka:2009un}. For a topological insulator, the gapless modes on the boundary are encoded in the degeneracies of the bulk ground state entanglement spectrum \cite{Fidkowski}. More generally, the concept of quantum entanglement plays important roles in characterizing the topological phase \cite{kitaev, wen}. Although the study in the previous subsection shows that the gapped system from holography in the presence of a boundary is a topologically trivial insulator, it should still be  interesting to explore its entanglement entropies. In this subsection, we study the  entanglement entropies of the gapped system with boundaries from AdS/BCFT. 

It is known that the entanglement entropy is dominated by the divergent area law with the UV cutoff. In the  presence of a boundary, additional terms might contribute to the entanglement entropy \cite{Fujita:2011, Seminara:2018pmr}. In \cite{Myers:2012ed, Liu:2012eea, Liu:2013una}, a renormalized entanglement entropy, which is finite and independent of the UV cutoff, has been introduced to characterize the entanglement at a certain length (or energy) scale. We will generalize it to the case of BCFT. 

We will first compute the entanglement entropy and then study the renormalized entanglement entropy for the gapped system in AdS/BCFT.   
The subsystem under consideration is an infinite strip adjacent to the boundary, 
i.e., $0<x<\ell$, while $y$ is infinite which will be renormalized to be $y\in [-L, L]$ with $L\to\infty$. The minimal surface $\gamma$ is specified by $u=u(x)$, which is a section at constant $y$. The extremal surface has the boundary condition $u(\ell)=0$.

The induced metric on $\gamma$ is
\be
ds^2_\gamma=\frac{1}{u^2}\bigg[\,\Big(1+\frac{u'^2}{f(u)}\Big)\,dx^2+dy^2\bigg]\,,
\ee
from which one obtains the area functional 
\be
\label{eq:areafun1}
A=2L\,\int_{x_*}^{\ell} dx\,\frac{1}{u^2}\sqrt{1+\frac{u'^2}{f}}\,.
\ee
When $f=1$, the above equations reduces to the  AdS$_4$/BCFT$_3$ in \cite{Seminara:2018pmr}. Here, we focus on the gapped geometries with $f$ shown in \eqref{eq:bgsol1}. 

\begin{figure}[h]
\begin{center}
\begin{tikzpicture}[scale=0.5]

\draw [blue!10] (0,0)--(7.5,0);
\draw [purple] (6,0)--(6,6) node[midway, right]{$\gamma_1$};
\draw [green, domain=-2.5:0] plot(\x,-sinh\x ); 

\draw [orange] (6,0) .. controls (6,1) and (5,4.5) .. (2,4.5) node[above]{$(x_t, u_t)$ } node[midway, left]{$\gamma_2$}.. controls (1,4.5) and (-1,4) .. (-1.5, 2.13) node[left]{$(x_*, u_*)$};
\draw[black] (-2.5, 6.0)  node[above]{ $Q $ };

\filldraw[orange] (2, 4.5) circle (2pt);
\filldraw[orange] (-1.5, 2.13) circle (2pt);
\filldraw[black] (0, 0) circle (2pt) node[below] {$O$};
\filldraw[black] (6, 0) circle (2pt) node[below] {$\ell$};

\draw[-latex, blue, opacity=0.8] (0, 0)--(8.5, 0) node[anchor=north, at end]{ $x$ };

\draw [blue!10] (15,0)--(17,0);
\draw [purple] (21,0)--(21,6) node[midway, right]{$\gamma_1$};

\draw [green, domain=15:17.7] plot(\x, -1155/2+227/2*\x -15/2* \x^2 + 1/6*\x^3); 

\draw [orange] (21,0) .. controls (21,1) and (19,3) .. (17, 10/3) node[midway, left]{$\gamma_2$} node[left]{$(x_*, u_*)$};

\draw[black] (17.7, 6.0)  node[above]{ $Q$ };
\filldraw[orange] (17, 10/3) circle (2pt);
\filldraw[black] (15,0) circle (2pt) node[below]{$O$};
\filldraw[black] (21, 0) circle (2pt) node[below]{$\ell$};

\draw[-latex, blue, opacity=0.8] (15, 0)--(23.5, 0) node[anchor=north, at end]{ $x$ };
\end{tikzpicture}
\end{center}
\vspace{-0.3cm}
\caption{\small Cartoon plot for the configuration of the extremal surfaces at $c\leq 0$ ({\em left}) and $c>0$ ({\em right}). The right plot is for the cases with $n>2$, while there is no configuration of $\gamma_1$ for $n=2$. We have suppressed the $y$ axis, and now the boundary theory lives along the $x$ axis. The Green line is the profile of the EOW brane $Q$ with $u(x)$ parametrized by \eqref{eq:bgeol2}. For the strip geometry we considered, there might exist two different kinds of extremal surfaces.}
\label{fig:exsurface0}
\end{figure}
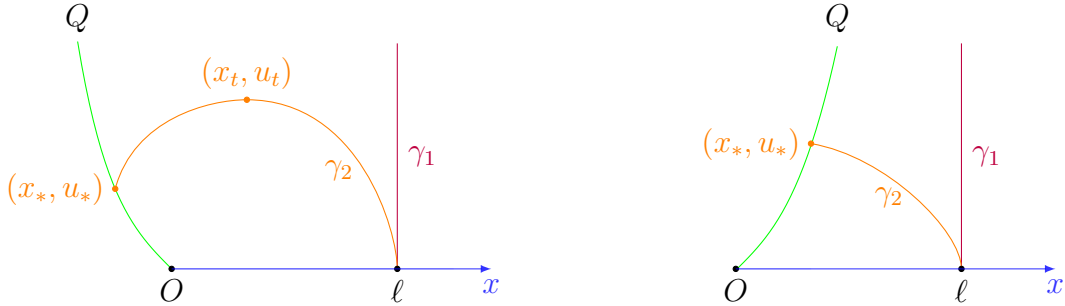

Since the above functional does not implicitly depend on $x$, there is a conserved quantity 
\be
\label{eq:consq}
\frac{1}{u^2\sqrt{1+\frac{u'^2}{f(u)}}}=C\,,
\ee
where $C$ is a constant. The final profile of the surface $\gamma$ depends on the value of $c$ which determines the embedding of the EOW brane $Q$ via \eqref{eq:bgeol2}. In the following, we will discuss the cases $c\leq 0$ and   
$c>0$ separately. For these two different cases, the cartoon plots of the extremal surfaces are shown in Fig. \ref{fig:exsurface0}.

\begin{itemize}
\item Case 1: $c\leq 0$
\end{itemize}

In the case $c<0$, the profile of the EOW brane $Q$ is along the regime $x\leq 0$ as shown in \eqref{eq:bgeol2}, while in the case $c=0$ the profile of $Q$ sits along the $u$ axis with $x=0$. Nevertheless, the properties of extremal surfaces in these two cases (except the case of $c=0, n=2$) share lots of similarities, and, therefore, we discuss them together.  

Intuitively, we expect to have two different kinds of local extrema of the area functional as shown in the left plot in Fig. \ref{fig:exsurface0}. One configuration is the surface $x=\ell$, which corresponds to $C=0$ in \eqref{eq:consq}, i.e., the purple line $\gamma_1$ in the left plot in Fig. \ref{fig:exsurface0}. 
This configuration exists at arbitrary value of $\ell>0$. As we will discuss later, for $\ell>\ell_c$, this is the unique configuration. 
The entanglement entropy is\footnote{Note that one can suppress $G$ using the unit $16\pi G=1$. In the following we will not do this.} 
\bea
\label{eq:ee1a}
\setlength\arraycolsep{1pt}
S=\frac{2L}{4G}\int_{u_c}^\infty  \frac{du}{u^2\sqrt{f}}\,
=\frac{2L}{4G} \frac{2}{(n+2) u_c^{(n+2)/2}}\, {}_2 F_1\left[\frac{1}{2},
\frac{1}{2}+\frac{1}{n},
\frac{3}{2}+\frac{1}{n}, 
- \frac{1}{u_c^n} \right] \,,
\eea
where $n\in[2,6]$, $G$ is the Newton constant, and $u_c$ is the cutoff near the boundary. When $u_c\to 0$, we have 
$u_c A/(2L) \to 1$. Note that the entanglement entropy \eqref{eq:ee1a} is independent of $\ell$. Therefore, we have $\partial{S}/\partial \ell=0$, which means that the renormalized entropy is zero for this configuration. 

Another kind of configuration is  shown as orange curved line $\gamma_2$ in the left plot in Fig. \ref{fig:exsurface0}. We have the turning point $(x_t, u_t)$ at which $u'(x_t)=0$ and the intersecting point $(x_*, u_*)$ between the extremal surface $\gamma_2$ and the EOW brane $Q$ where $n_Q\cdot n_\gamma=0$, i.e. 
$u'(x_*)=-c\sqrt{f}$.  
From \eqref{eq:consq}, we, therefore, have $C^{-1}=u_t^2=u_*^2\sqrt{1+c^2}$ which leads to 
\be \label{eq:eerel1}
u_t=u_* (1+c^2)^{1/4}
\ee
and 
\be
\label{eq:eerel2}
u'^2=\left(\,\frac{u_t^4}{u^4}-1\right)f\,.
\ee
When $c=0$, we have $u_t=u_*$. 
Note that for $x<x_t$ we have $u'>0$, while for $x>x_t$ we have $u'<0$, where the prime is the derivative with respect to $x$.

From \eqref{eq:eerel2}, we have the relation 
\be
\label{eq:eeeqn1}
\ell-x_*=\int_{u_*}^{u_t} du \frac{1}{\sqrt{(\frac{u_t^4}{u^4}-1\big)f}}+\int^{u_t}_0 du \frac{1}{\sqrt{(\frac{u_t^4}{u^4}-1\big)f}}\,.
\ee
Note that $u_*(x_*)$ is given by \eqref{eq:bgeol2}. 
From (\ref{eq:eeeqn1}, \ref{eq:eerel1}) and the relation  \eqref{eq:bgeol2} which relates $u_*(x_*)$, one could obtain $u_t$ as a function of $\ell$ as $ u_t=u_t(\ell, c, n)$. 

\begin{figure}[h]
\begin{center}
\includegraphics[width=0.45\textwidth]{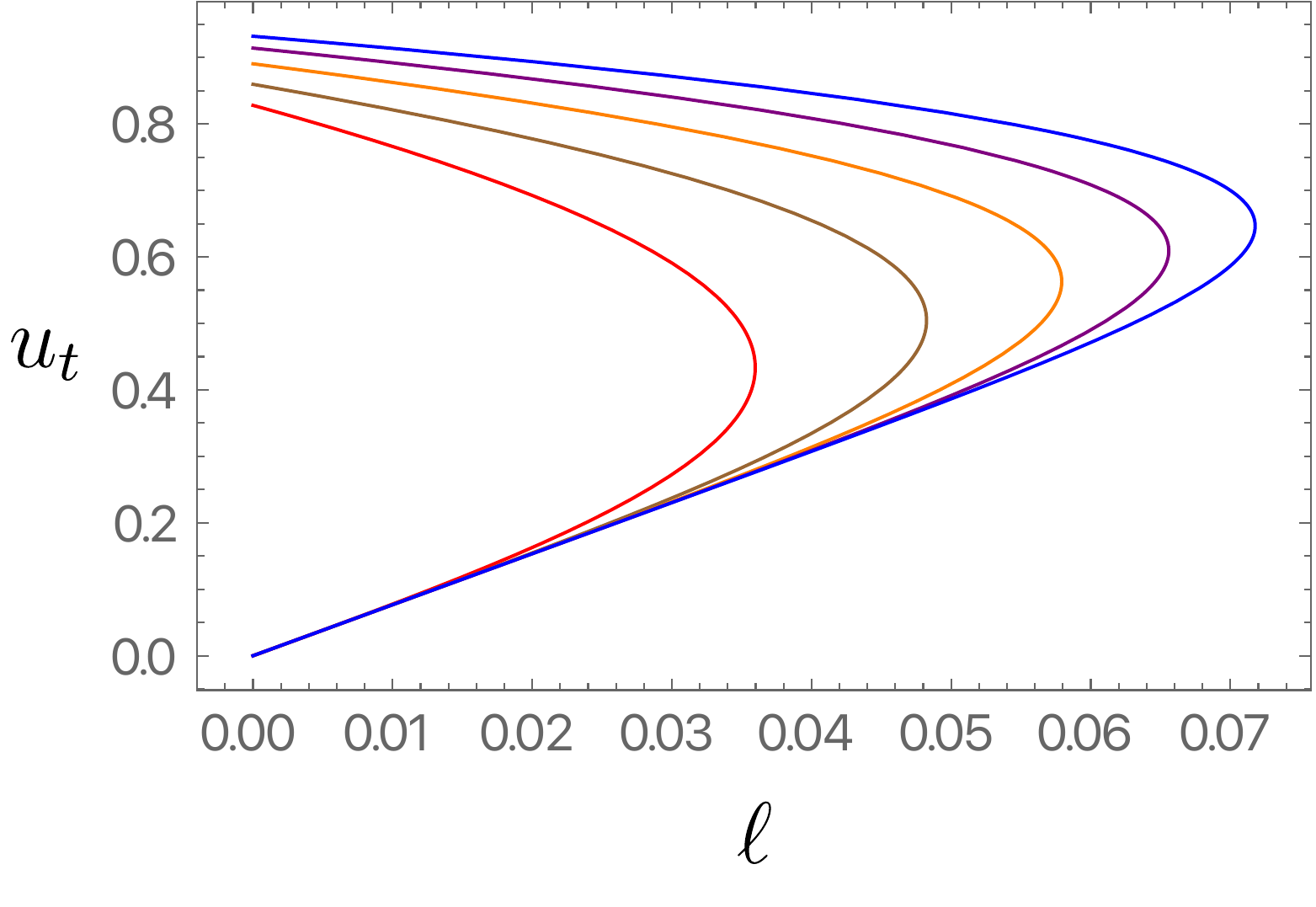}
~~~
\includegraphics[width=0.46\textwidth]{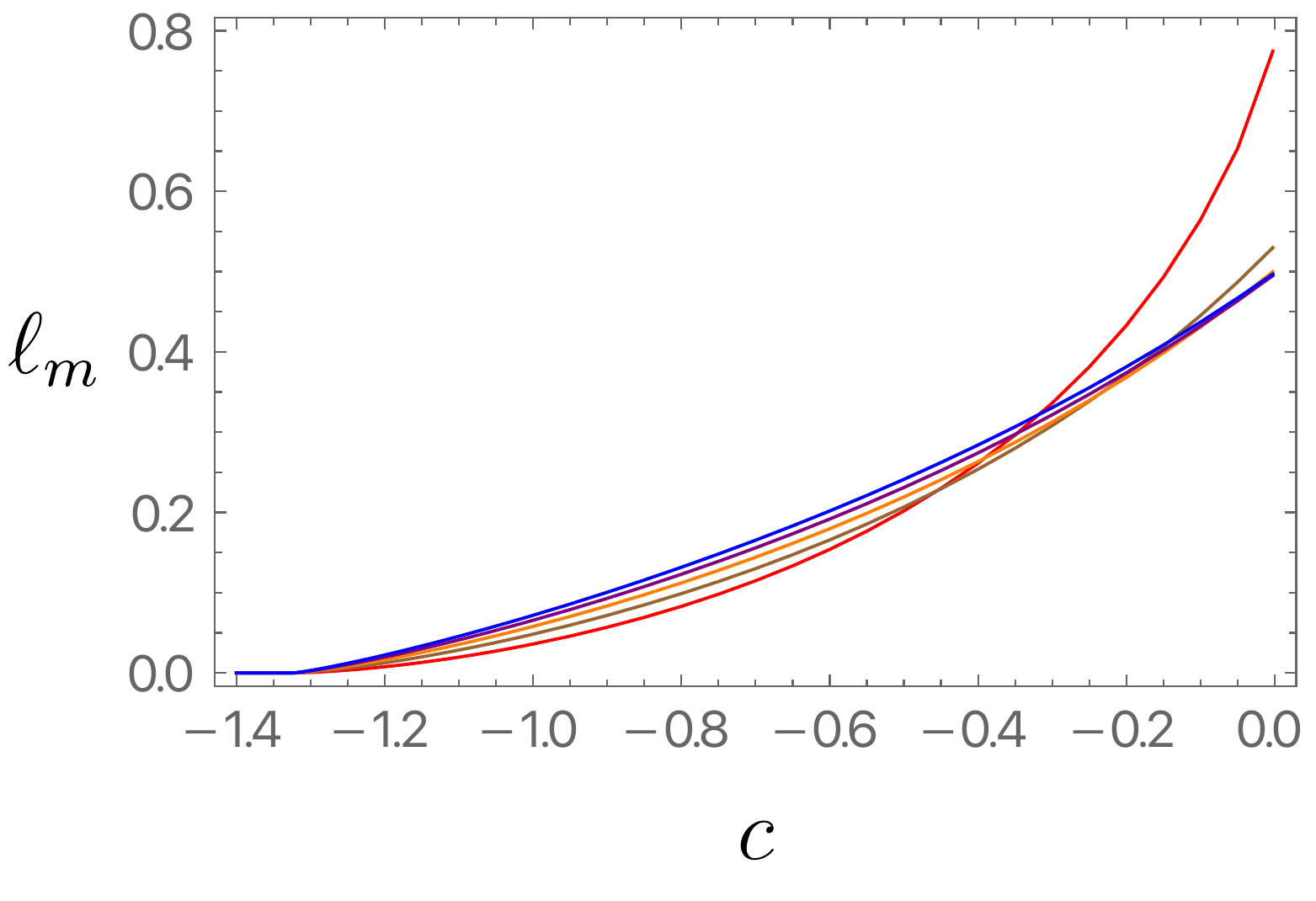}
\end{center}
\vspace{-0.3cm}
\caption{\small {\em Left: }  the location of the turning point $u_t$ as a  function of the width of the strip $\ell$ when $c=-1$ and different $n$. {\em Right}: the maximal value of the width $\ell_m$ below which there are two extremal curved surfaces as a function of the tension parameter $c$ for different $n$ when $c\leq 0$. There exists a critical value of $c\approx -1.32$
below which we do not have a curved extremal surface. 
In these two plots, we have $n=2$ (red),$\,3$ (brown), $\,4$ (orange), $\,5$ (purple), and $6$ (blue).  
}
\label{fig:config2}
\end{figure}

Equation \eqref{eq:eeeqn1} can be solved only numerically. In the left plot in Fig. \ref{fig:config2}, we show the dependence of $u_t$ as a function of $\ell$ for different $n$. We can see the existence of a maximal $\ell_m$. Below it, i.e., $\ell<\ell_m$, there exist two different extremal surfaces in addition to the configuration of the straightforward line. Above $\ell_m$, the configuration of this kind (i.e., the orange curve in the left plot in Fig. \ref{fig:exsurface0}) does not exist. This is different from the pure AdS case with a negative tension on the EOW brane, in which there does not exist a maximal $\ell$ which separates the topology of minimal surfaces \cite{Seminara:2018pmr}.   
In the right plot in Fig. \ref{fig:config2}, we show the dependence of $\ell_m$ as a function of $c$ for different $n$. We found that $\ell_m$ decreases when $c$ becomes smaller and there exists a critical value $c_m$ such that below it we do not have any curved configuration of extremal surface. This reminds us of the existence of the critical tension for the extremal surfaces in the pure  AdS/BCFT \cite{Seminara:2018pmr}.

\begin{figure}[h!]
\begin{center}
\includegraphics[width=0.62\textwidth]{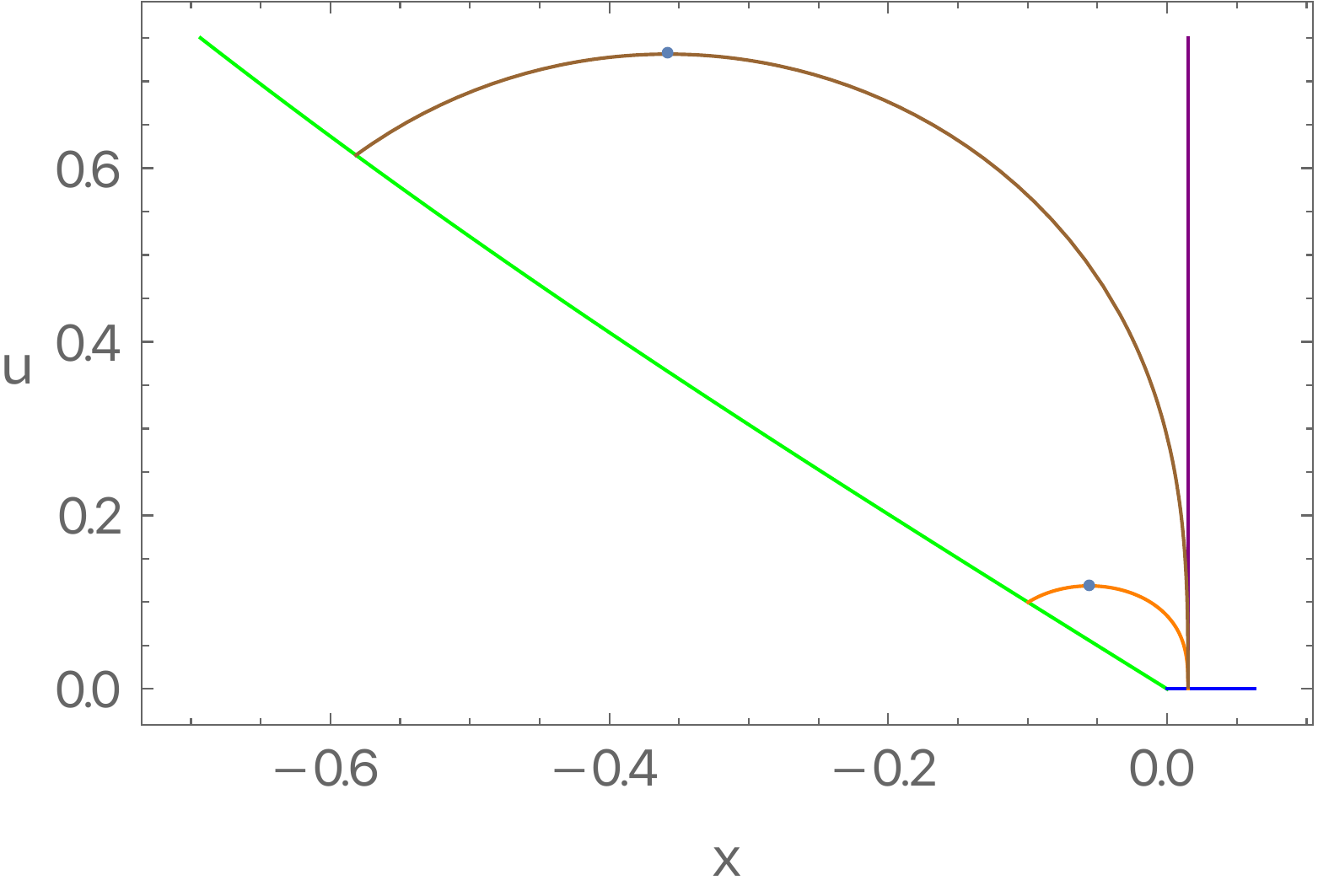}
\end{center}
\vspace{-0.4cm}
\caption{\small Plot of the extremal surfaces for $c=-1,\, n =2$. In this case, $\ell_m\approx 0.036$. When $\ell=0.015<\ell_m$, there exist three different extremal surfaces, including the curved brown and orange lines and the straight purple line. The dots on the curves are the locations of the turning points. 
}
\label{fig:es}
\end{figure}

Fig. \ref{fig:es} shows examples of extremal surfaces for a specific value of $n=2, c=-1$. The BCFT lives in $x\geq 0$, and the green line refers to the location of the EOW brane $Q$. We choose one specific value of $\ell$ with $\ell<\ell_m$ and plot the two curved (brown and orange)  and one straight (purple) extremal surfaces. For any value of $\ell$, the vertical line of extremal surface always exists. When $\ell$ is smaller than $\ell_m$, there exist three different configurations of extremal surfaces. When $
\ell=\ell_m$, there exist two different configurations of extremal surface  where the two curved lines merge into the same line, while when $\ell$ is greater than $\ell_m$, there exists only one extremal surface which is the straight line.

\begin{figure}[h!]
\begin{center}
\includegraphics[width=0.49\textwidth]{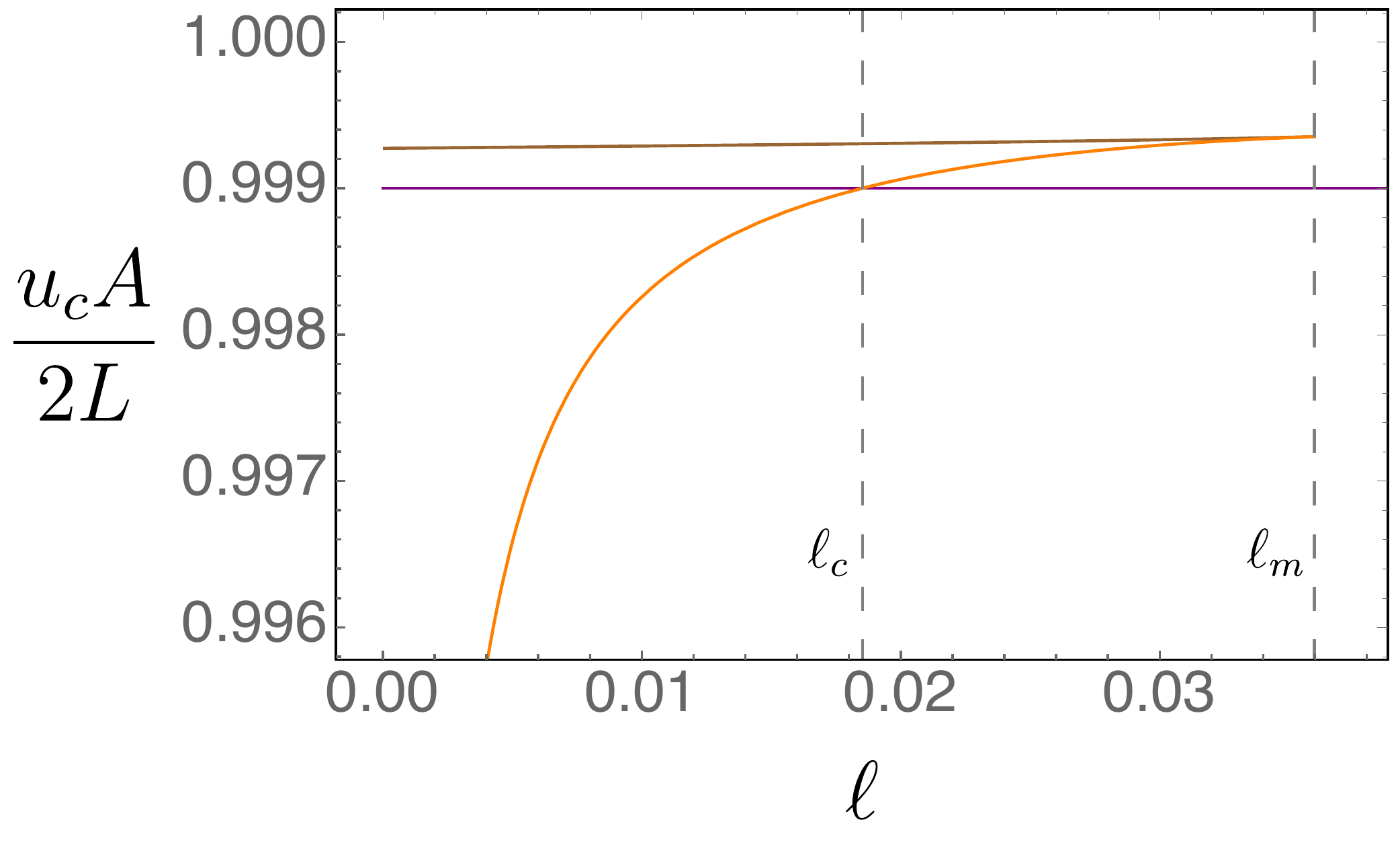}
~~~
\includegraphics[width=0.46\textwidth]{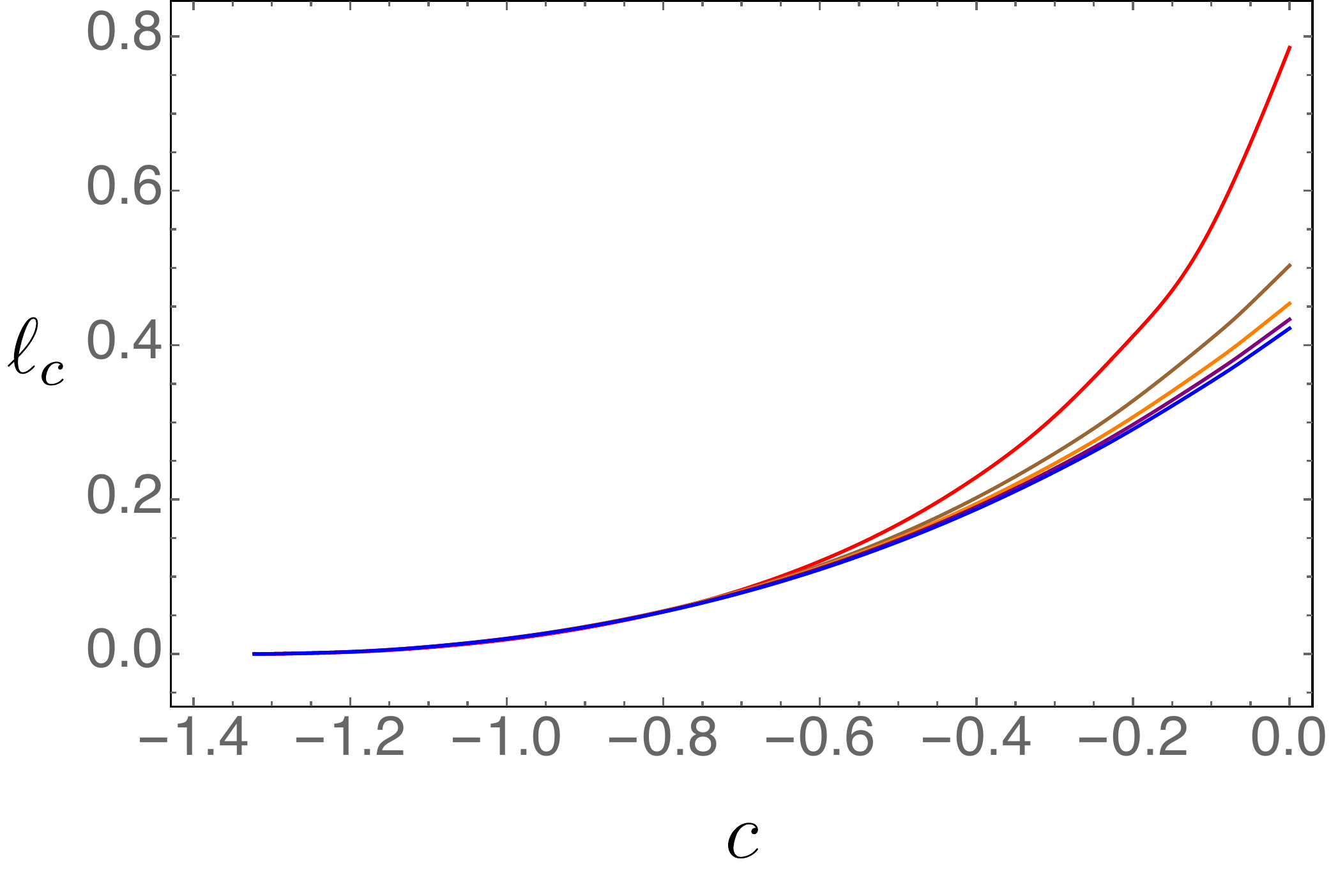}
\end{center}
\vspace{-0.4cm}
\caption{\small {\em Left:} plot of entanglement entropy $\frac{u_cA}{2L}$ as a function of $\ell$ for $c=-1, n=2$. We have used $u_c=10^{-3}$. Note that, when $\ell$ is small, one needs to choose smaller $u_c$ to make sure $\epsilon\ll \ell$. 
{\em Right:} the critical length $\ell_c$ as a function of $c$ for different $n$.}
\label{fig:eec>0}
\end{figure}

The entanglement entropy can be obtained from the  area of the extremal surfaces 
\be
\label{eq:ee1}
\begin{split}
S&= \frac{A}{4G}=\frac{L}{2G} \,\int_{x_*}^{\ell-\epsilon} dx\,\frac{1}{u^2}\sqrt{1+\frac{u'^2}{f}}\,.
\end{split} 
\ee
As seen from the discussions above, there might be multiple extremal surfaces and the RT surface is determined by the one with minimal area.  Using the same cutoff $u_c$ which satisfies $u(\ell-\epsilon)=u_c$, the areas of the extremal surfaces $ \frac{u_c A}{2L}$ for $n=2, c=-1$ are shown in Fig. \ref{fig:eec>0}. We find that there exists a critical $\ell_c$ below which the orange curve has minimal area, while above $\ell_c$ the straight vertical purple line has minimal area. There is a first-order transition at $\ell_c$. Moreover, we always have  $\ell_c<\ell_m$. These phenomena are quite general for any $c\leq 0$ except the case of $c=0, n=2$, which we will comment on at the end of this part.

We find that the entanglement entropy satisfies
\bea
S=\frac{A}{4G}=\frac{L}{2G}\,\Big[ \frac{a_1}{u_c}-a_2+\mathcal{O}(u_c) \Big] \,,
\eea
where $a_1$ is approximately equal to $1$, which can be seen from \eqref{eq:ee1}. Furthermore, $a_2 ~(a_2>0)$ is a function of $\ell$ and independent on the cutoff. From \eqref{eq:areafun1}, when $\ell\to 0$, we have $f\to 1$, $u'\to -\infty$, and one expects $a_2\propto 1/\ell$ at very small $\ell$. For larger $\ell>\ell_c$, from \eqref{eq:ee1a}  we have $a_2=0$. 
For other values of $\ell$, 
we have to obtain the behavior of $a_2$ numerically. 
For holographic CFTs without a boundary, we have $a_2\propto 1/\ell$ with a constant coefficient \cite{Nishioka:2009un}, while for AdS$_4$ plus the EOW  brane with constant tension, we also have $a_2\propto 1/\ell$ with the coefficient depending on the effective tension of the EOW brane \cite{Seminara:2018pmr}.
Nonetheless, in our case, $a_2$ has a complicated dependence on $\ell$.

\begin{figure}[h]
\begin{center}
\includegraphics[width=0.65\textwidth]{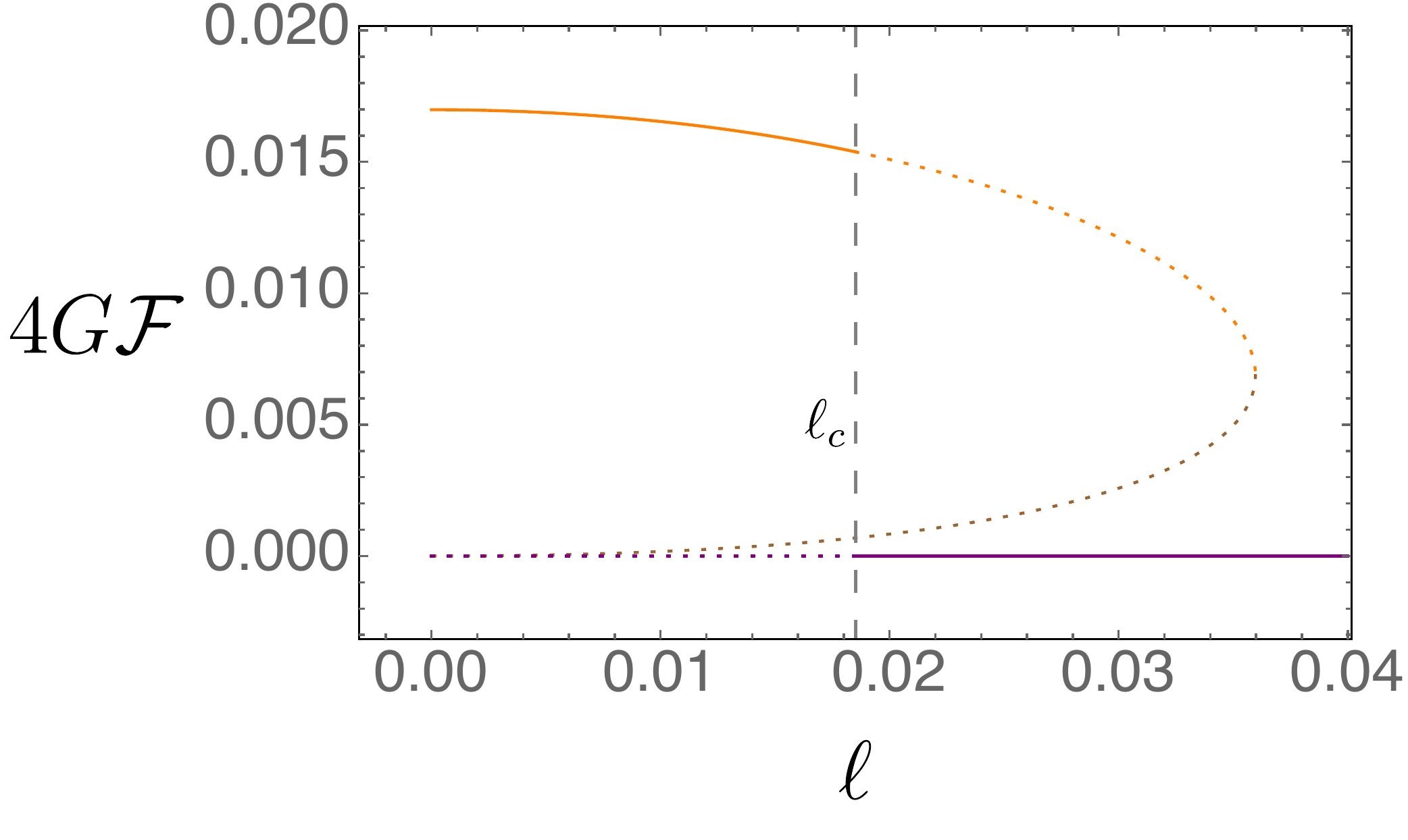}
\end{center}
\vspace{-0.4cm}
\caption{\small Plot of the renormalized entanglement entropy $4G \mathcal F$ as a function of $\ell$ when $c=-1$ and $n=2$. The solid lines are for the minimal surfaces, while the dashed brown, yellow, and purple lines are for the nonminimal extremal surfaces. The dashed black line is the location of the transition $\ell_c$.
}
\label{fig:reeF}
\end{figure}

Now let us discuss the renormalized entanglement entropy following \cite{Myers:2012ed,Liu:2013una}. Close to $u\to 0$, from \eqref{eq:eerel2} we have 
\be\label{eq:uto01}
x(u)=\ell-\frac{u^3}{3 u_t^2}+\cdots\,.
\ee
From the variation of \eqref{eq:ee1} with respect to $\ell$ and using \eqref{eq:uto01}, we have 
\bea 
\label{eq:effent}
\mathcal{F}=
 \frac{\ell^2}{2L}\frac{\partial S}{\partial \ell } =\frac{1}{4G}\frac{\ell^2}{u_t^2}\,.
\eea
The detailed derivation of the above equation can be found in appendix \ref{app:ree}. 
We see that the renormalized entanglement entropy $\mathcal{F}$ is determined by $\ell$ and $u_t$, 
which takes the similar form as the case of AdS$_4$/CFT$_3$ without a boundary \cite{Myers:2012ed,Liu:2013una}.  
However, now the detailed dependence of $u_t$ on $\ell$ is different from the case without a boundary. Compared with AdS$_4$/BCFT$_3$ in pure AdS$_4$, where $\mathcal{F}$ is independent of $\ell$ \cite{Chu:2017aab,Seminara:2018pmr},  now we have interesting nontrivial 
$\ell$ dependence of $\mathcal{F}$ as shown in Fig. \ref{fig:reeF}: When $\ell<\ell_c$, $\mathcal{F}$ is positive and monotonically decreasing; at  $\ell=\ell_c$,  there is a discontinuity for $\mathcal{F}$; and when $\ell>\ell_c$, $\mathcal{F}=0$. 

\begin{figure}[h!]
\begin{center}
\includegraphics[width=0.46\textwidth]{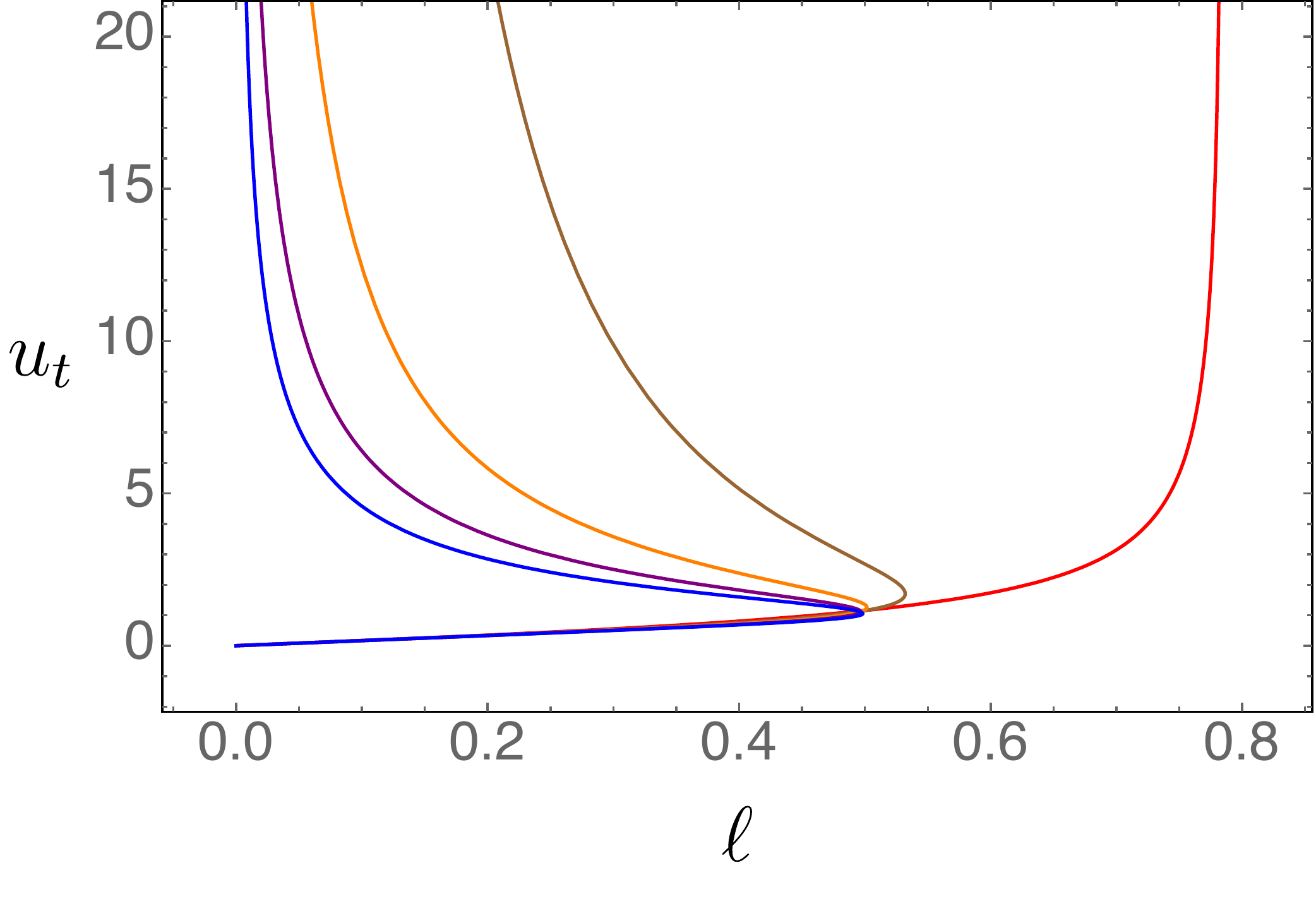}
~~~
\includegraphics[width=0.45\textwidth]{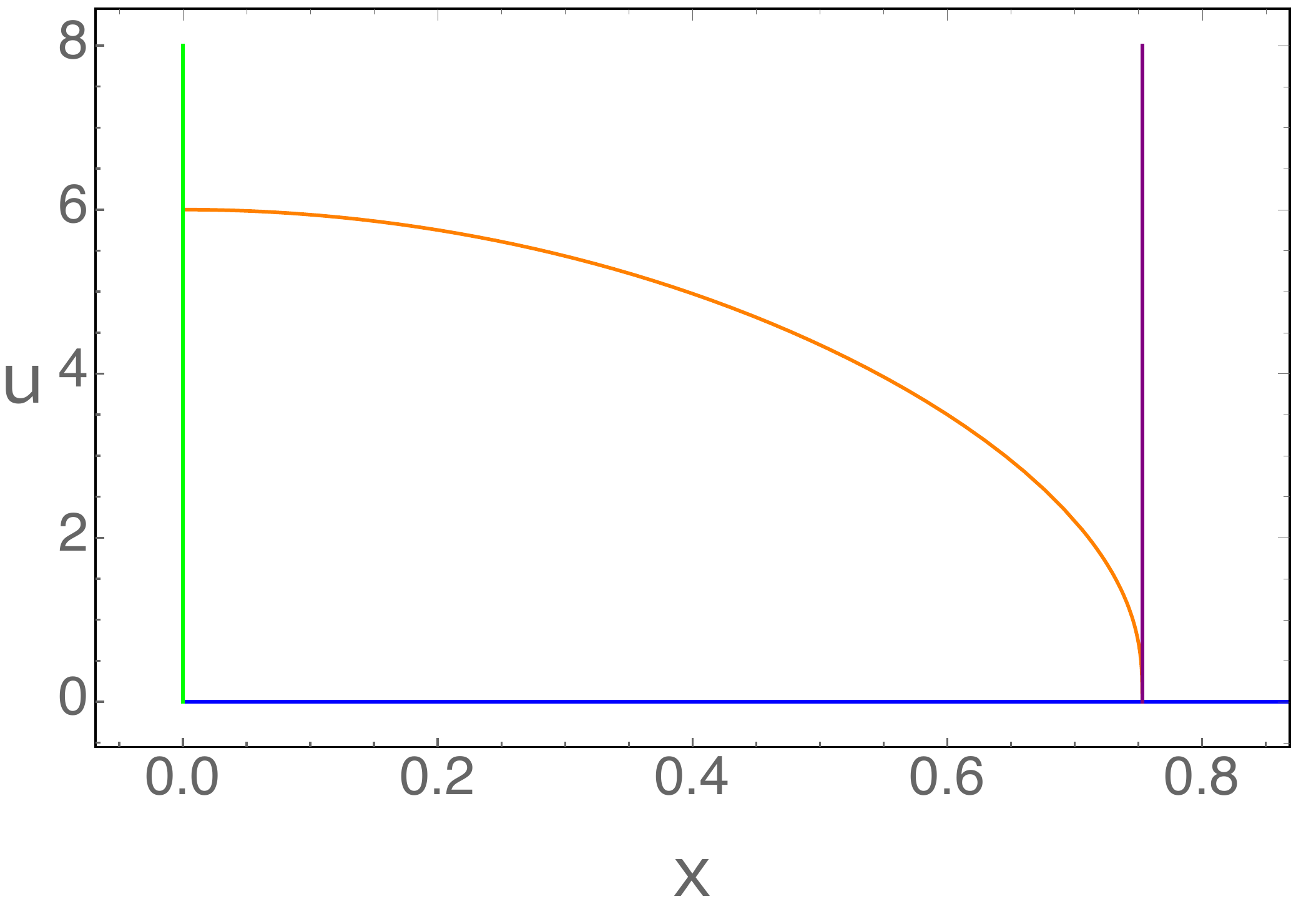}
\end{center}
\vspace{-0.3cm}
\caption{\small  {\em Left:} plots of $u_t$ as functions of $\ell$ when $c=0$ and $n=2$ (red),$\,3$ (brown), $\,4$ (orange), $\,5$ (purple), and $6$ (blue). {\em Right:} the extremal surfaces for $c=0, n=2$. In this case, $\ell_m\simeq 0.785$, and we have chosen $\ell=0.753$ (with $\ell<\ell_m$) so that there exist two different extremal surfaces.
}
\label{fig:config5}
\end{figure}

We make some comments on the case of $c=0, n=2$. In this case, the extremal surface behaves differently comparing to other cases of $c\leq 0$ (the left plot in Fig. \ref{fig:config5}); i.e., when $\ell<\ell_m$, there is only one curved extremal surface in addition to the straight vertical one, while in other cases there exist two different curved extremal surfaces. In the right plot in Fig. \ref{fig:config5}, we show one example of the extremal surface for $\ell<\ell_m$. When $\ell>\ell_m$, we have only one single straight extremal surface, which is the same as other cases. 

\begin{figure}[h!]
\begin{center}
\includegraphics[width=0.46\textwidth]{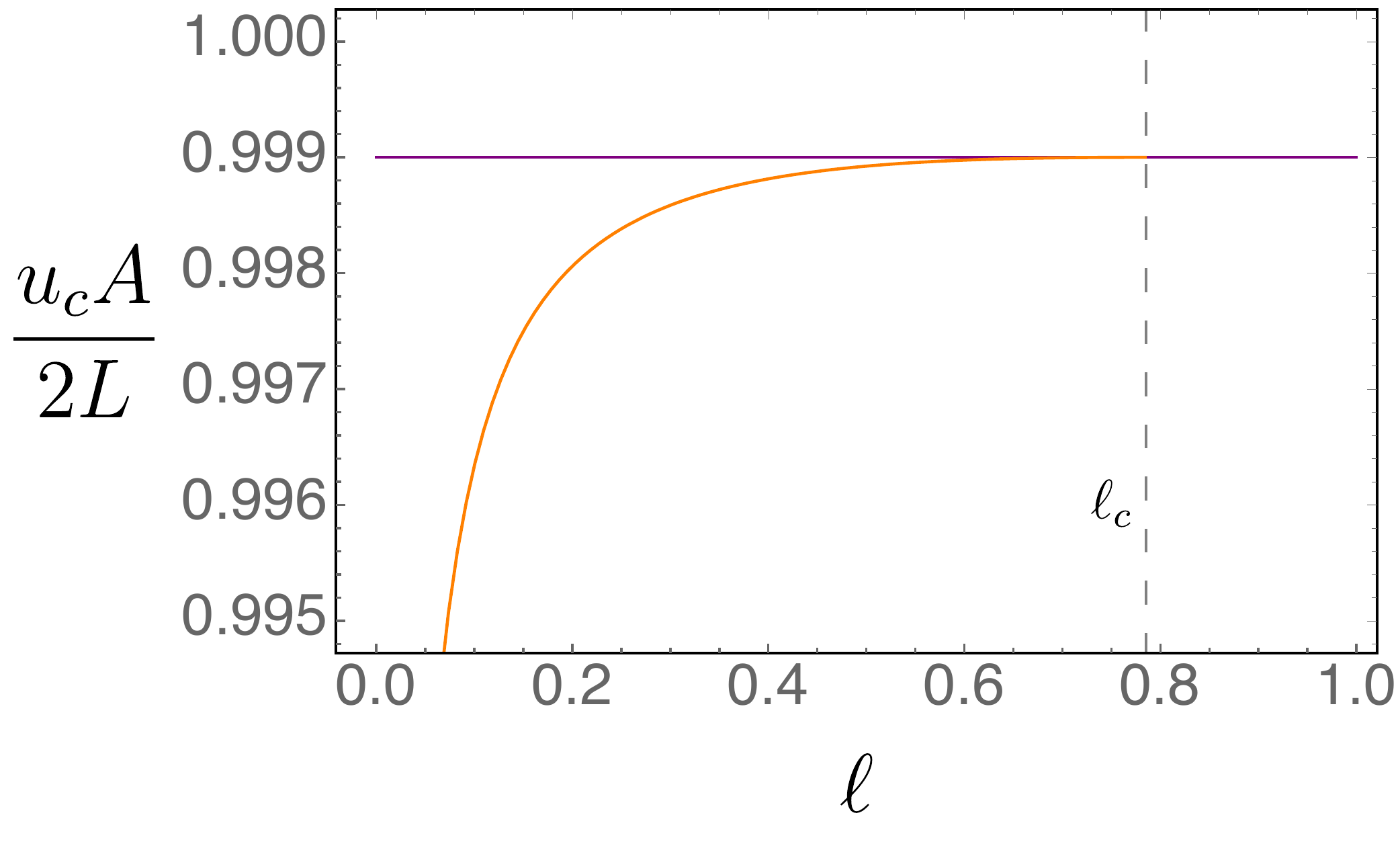}
\includegraphics[width=0.45\textwidth]{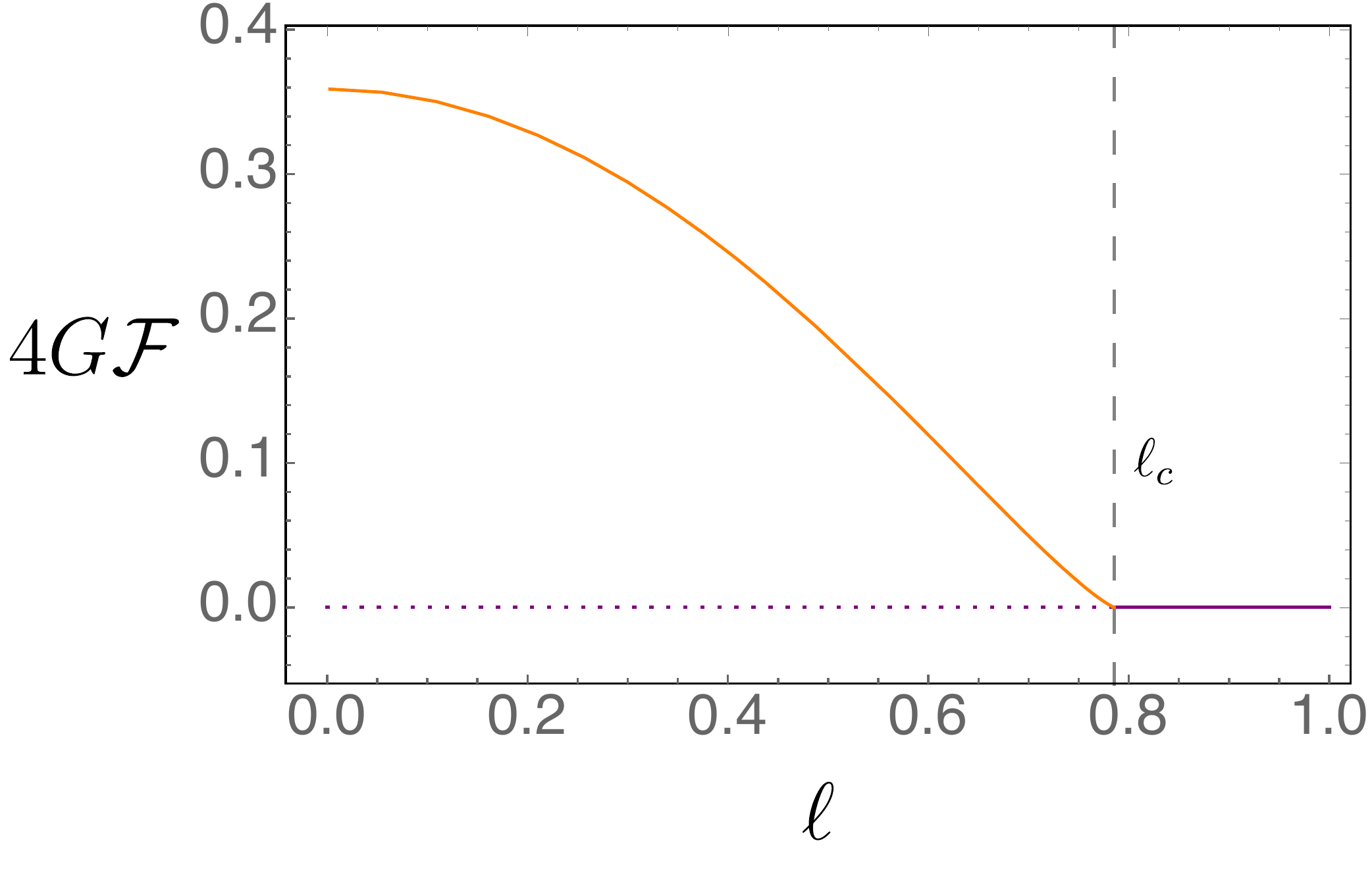}
\end{center}
\vspace{-0.3cm}
\caption{\small {\em Left:} plots of the area of the extremal surfaces as a function of $\ell$ when $c=0, n=2$. We set the cutoff $u_c=10^{-3}$.
{\em Right:} the renormalized entanglement entropy $4G \mathcal F$ as a function of $\ell$ for $c=0, n=2$. 
}
\label{fig:ee0n2}
\end{figure}

Fig. \ref{fig:ee0n2} shows the area of the extremal surfaces (left plot) and the renormalized entanglement entropy (right plot) as a function of the width of the strip $\ell$. We see that different from the other cases discussed in this part (e.g., Fig. \ref{fig:eec>0}), the area of the curved extremal surface is equal to the area of the straight vertical surface at $\ell=\ell_m$. The renormalized entanglement entropy is positive and monotonically decreasing when $\ell<\ell_c$ and is continuous while not smooth at $\ell=\ell_c$,  which is different from the other cases of $c\leq 0$. 

Comparing the right plot in Fig. \ref{fig:ee0n2} with the one in Fig. \ref{fig:reeF}, we see the renormalized entanglement entropy $\mathcal{F}$ behaves differently. Note that, in both these two cases, we have fixed the same metric in $N$ (i.e., $n=2$) while different values of $c$, which plays the role of the effective tension of the EOW brane. This indicates that the boundary of BCFT has nontrivial effects on the renormalized entanglement entropy, i.e., the number of the effective degrees 
of freedom inside the strip. 

\begin{itemize}
\item Case 2: $c > 0$
\end{itemize}

In this case, we have different profiles of the EOW branes, which depends on the value of $n$. As can be seen from Fig. \ref{fig:config}, when $n=2$ the EOW brane will approach $x\to\infty$, while when $n\in (2, 6]$ the EOW brane can approach only a finite value of $x_m$.  
Thus, in the case $n=2$,  there is only one kind of extremal surface, while in the latter case there might be two different kinds of extremal surfaces when $x>x_m$, as shown in the right plot in Fig. \ref{fig:exsurface0}. In the following, we will study these two cases separately. 

The entanglement entropy associated with the straight line has the same form as \eqref{eq:ee1a}. We focus on the configuration of the curved extremal surface, i.e., the curved line $\gamma_2$ in the right plot in Fig. \ref{fig:exsurface0}. 
The intersecting point between the extremal surface and the EOW brane $(x_*, u_*)$ satisfies $n_Q\cdot n_\gamma=0$, i.e.,  
$u'(x_*)=-c\sqrt{f}$. Then $C^{-1}=u_*^2\sqrt{1+c^2}$. 
Therefore, we have 
\be
\label{eq:eerel3}
u'=-\sqrt{f}\,\bigg(\frac{u_*^4}{u^4}(1+c^2)-1\bigg)^{1/2}\,.
\ee

\begin{figure}[h!]
\begin{center}
\includegraphics[width=0.44\textwidth]{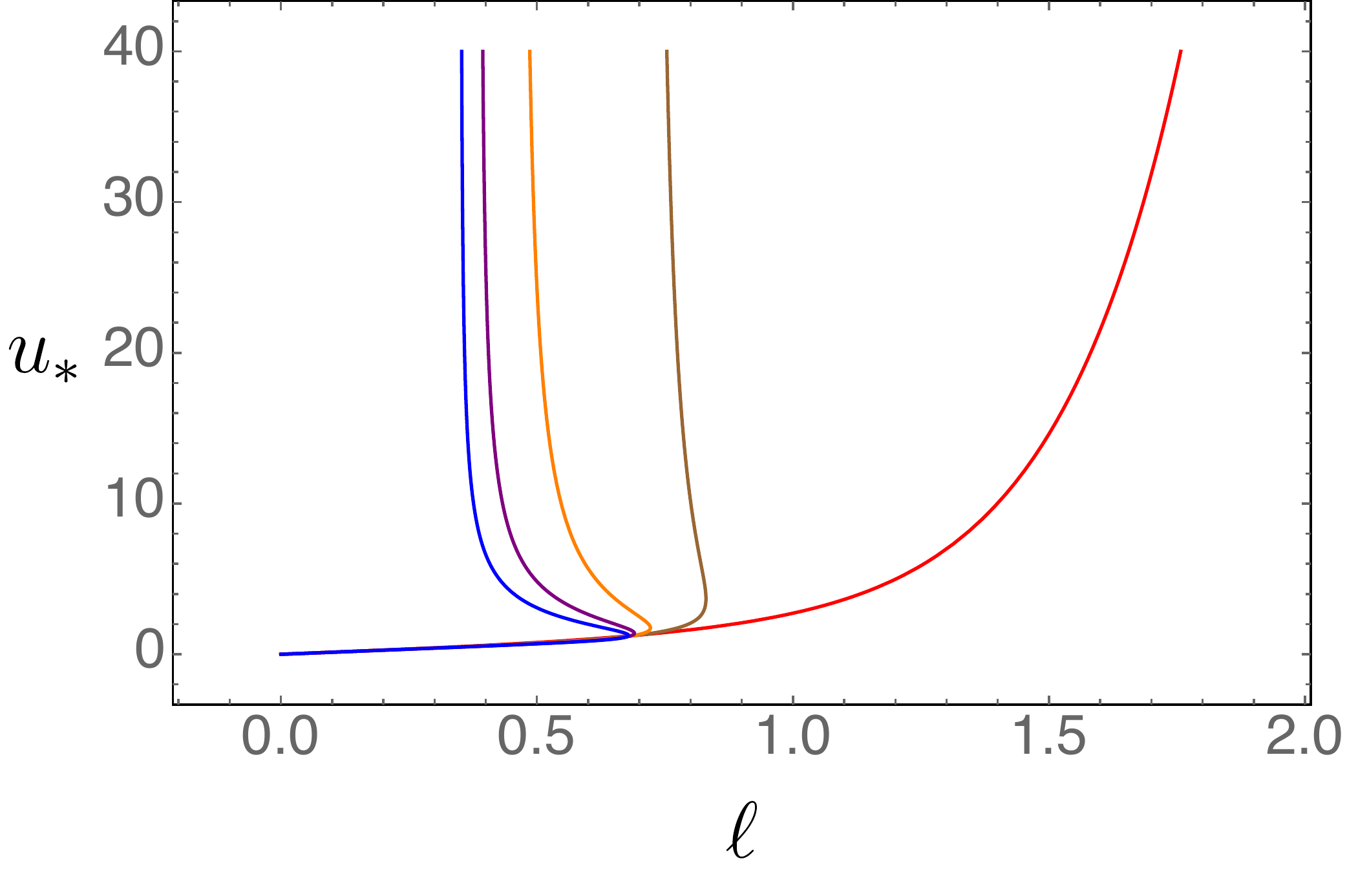}
~~
\includegraphics[width=0.44\textwidth]{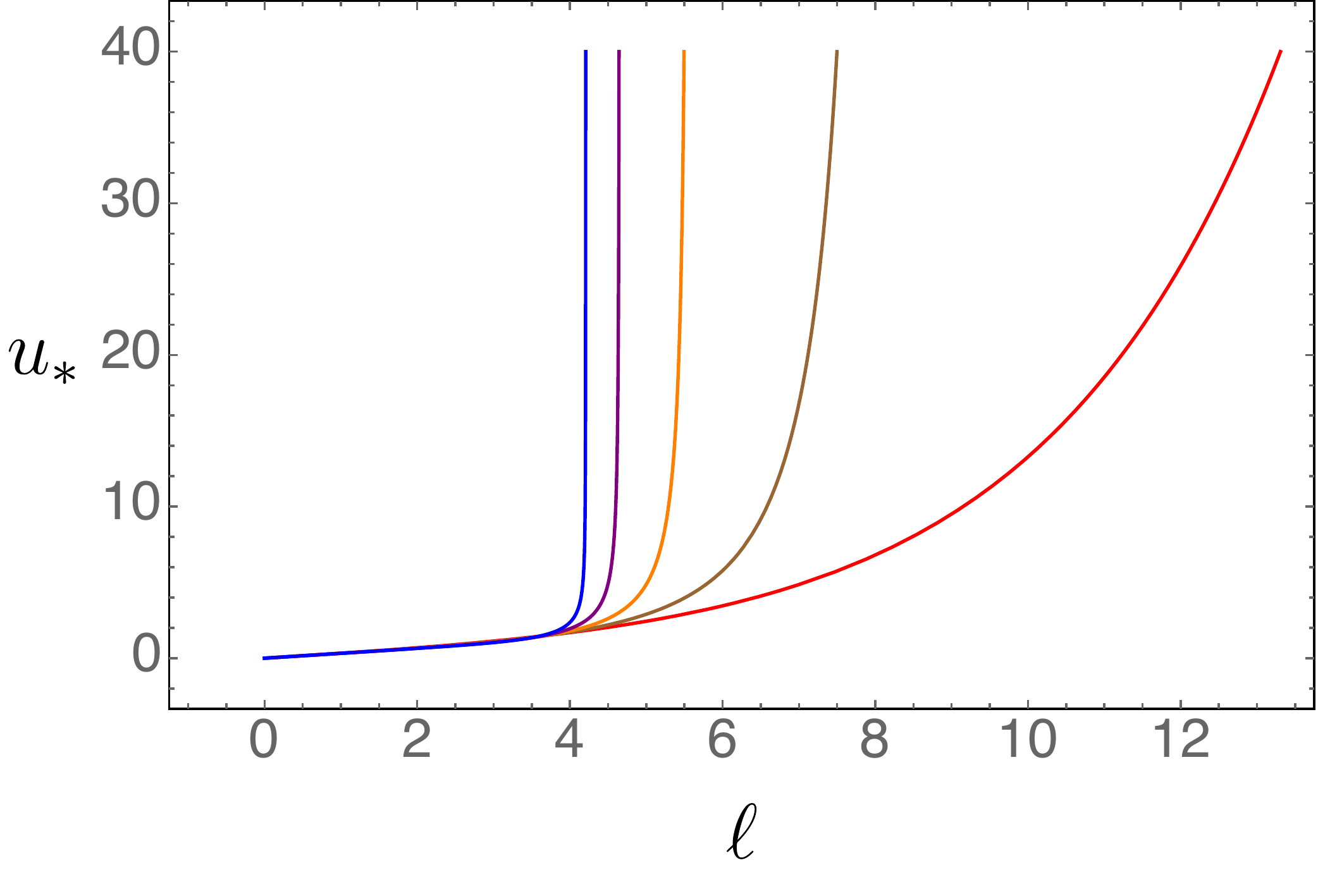}
\end{center}
\vspace{-0.3cm}
\caption{\small The location of the  interacting point between the extremal surface and the EOW brane $u_*$ as a function of $\ell$ for $c=1/4$ ({\em left}) and $c=3$ ({\em right}). In both cases, we have $n=2$ (red),$\,3$ (brown), $\,4$ (orange), $\,5$ (purple), and $6$ (blue).
}
\label{fig:cf-sol}
\end{figure}

Then we have the relation 
\begin{align}
    \ell-x_*=-\int_{u_*}^0 du\, \frac{1}{\sqrt{f}\,\bigg(\frac{u_*^4}{u^4}(1+c^2)-1\bigg)^{1/2}}\,.
\end{align}
As we know $x_*(u_*)$ from equation \eqref{eq:bgeol2} for the EOW brane $Q$, one can obtain $u_*$ as a function of $c, n$ and $\ell$. The plots for $u_*$ as functions of $\ell$ at two different values of $c$ and different $n$'s are shown in Fig. \ref{fig:cf-sol}. For $n=2$, we see that $u_*$ is monotonically increasing when we increase $\ell$, and this is what we expected, because the EOW brane approaches $x\to\infty$. For other values of $n=3,4,5,6$, there exists a critical value of $c$ which separates different behaviors of the extremal surfaces. In the left plot with $c<c_m$, we find that, when $\ell<x_m$, there exists only one curved extremal surface, while when $x_m<\ell<\ell_m$, there exist two different curved surfaces, and when $\ell_m<\ell$, there does not exist any curved extremal surface. Note that the vertical straight extremal surface shown as $\gamma_1$ in the right plot in Fig. \ref{fig:exsurface0} exits when $x_m<\ell$. When we increase $c$ to make it larger than a critical value $c_m$, as shown in the right plot in Fig. \ref{fig:cf-sol}, we find that when $\ell<x_m$ there exists only one curved extremal surface, while when $\ell>x_m$ there exists only the vertical straight extremal surface. 

\begin{figure}[h!]
\begin{center}
\includegraphics[width=0.32\textwidth]{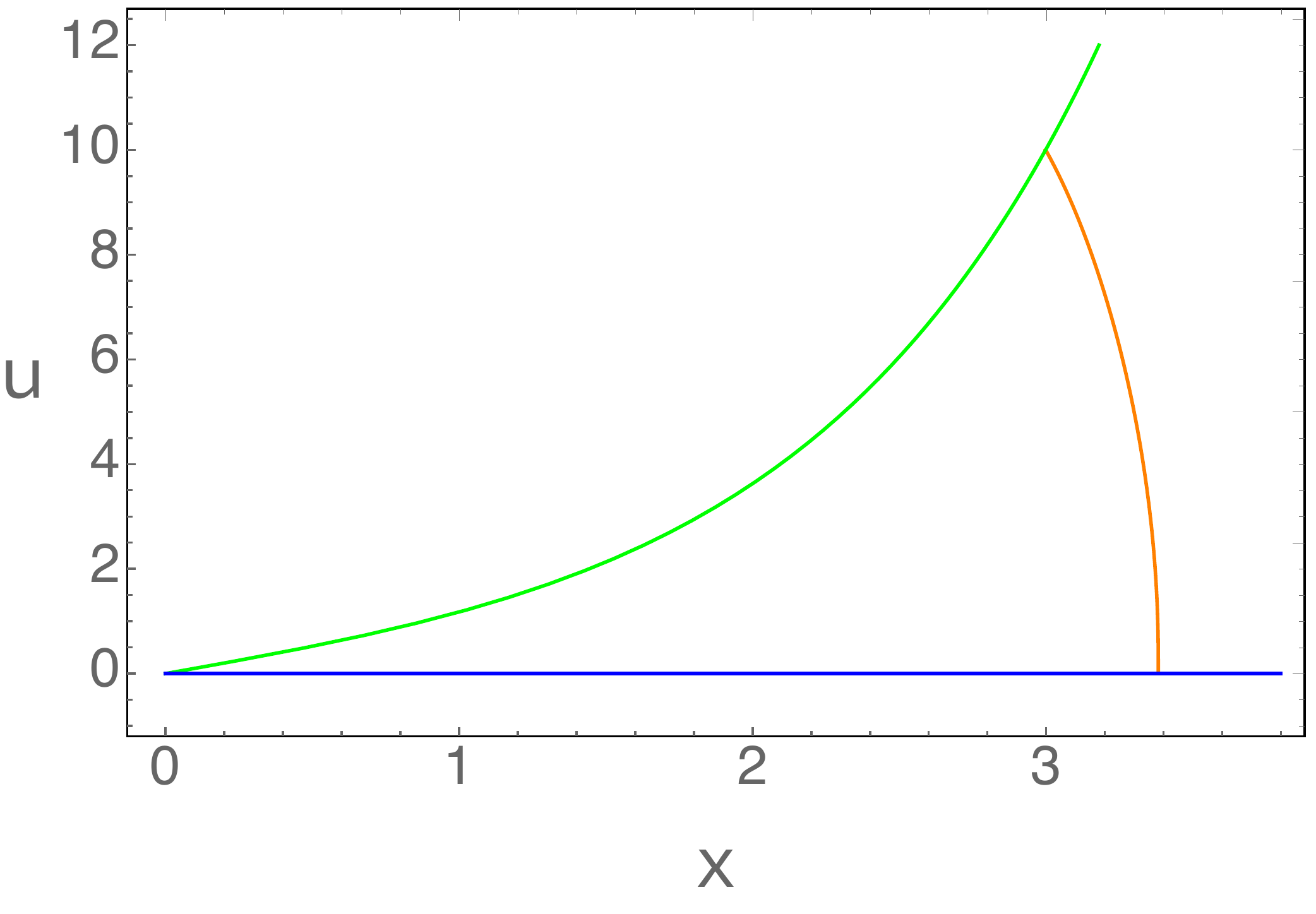}
~
\includegraphics[width=0.31\textwidth]{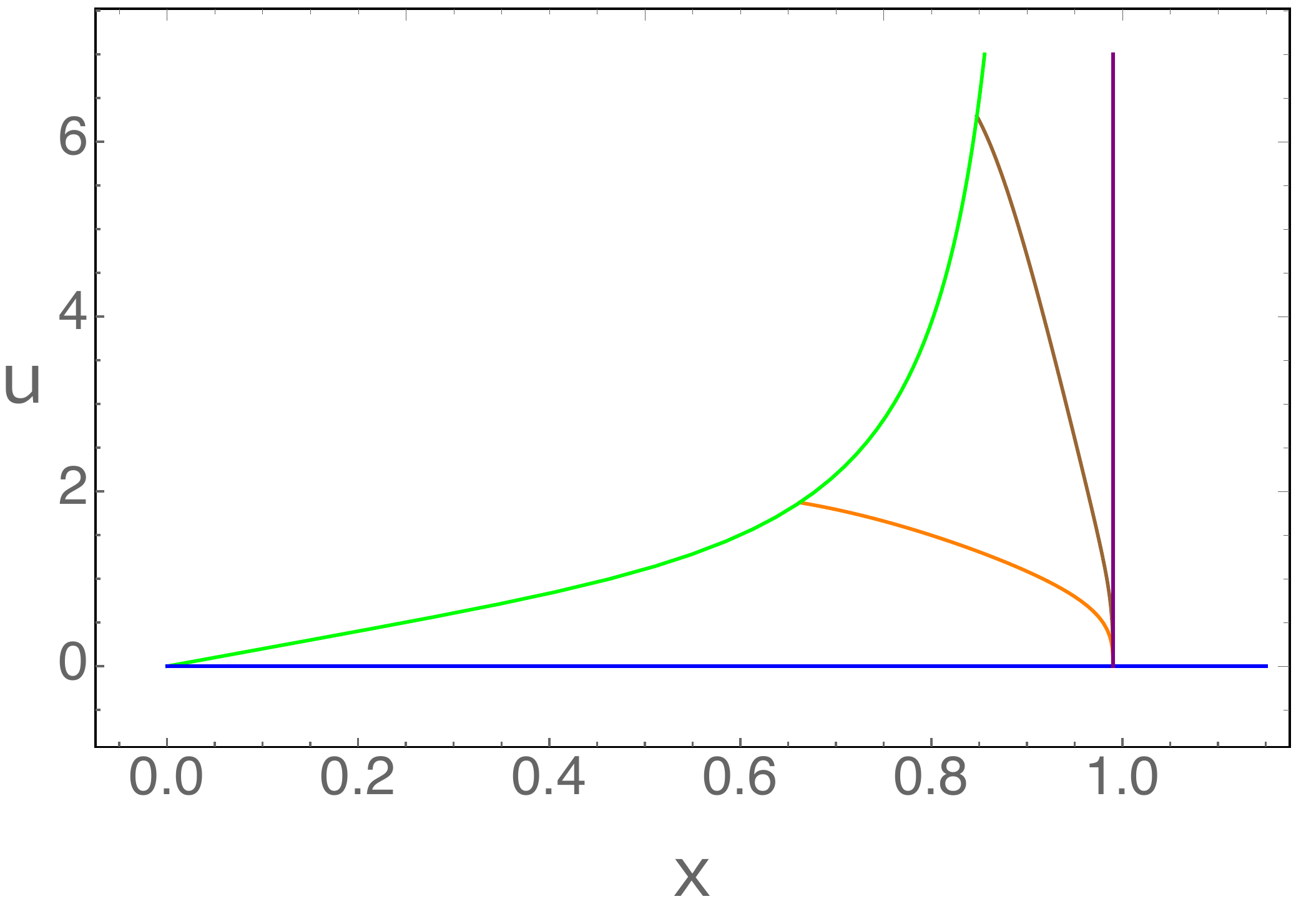}
~
\includegraphics[width=0.32\textwidth]{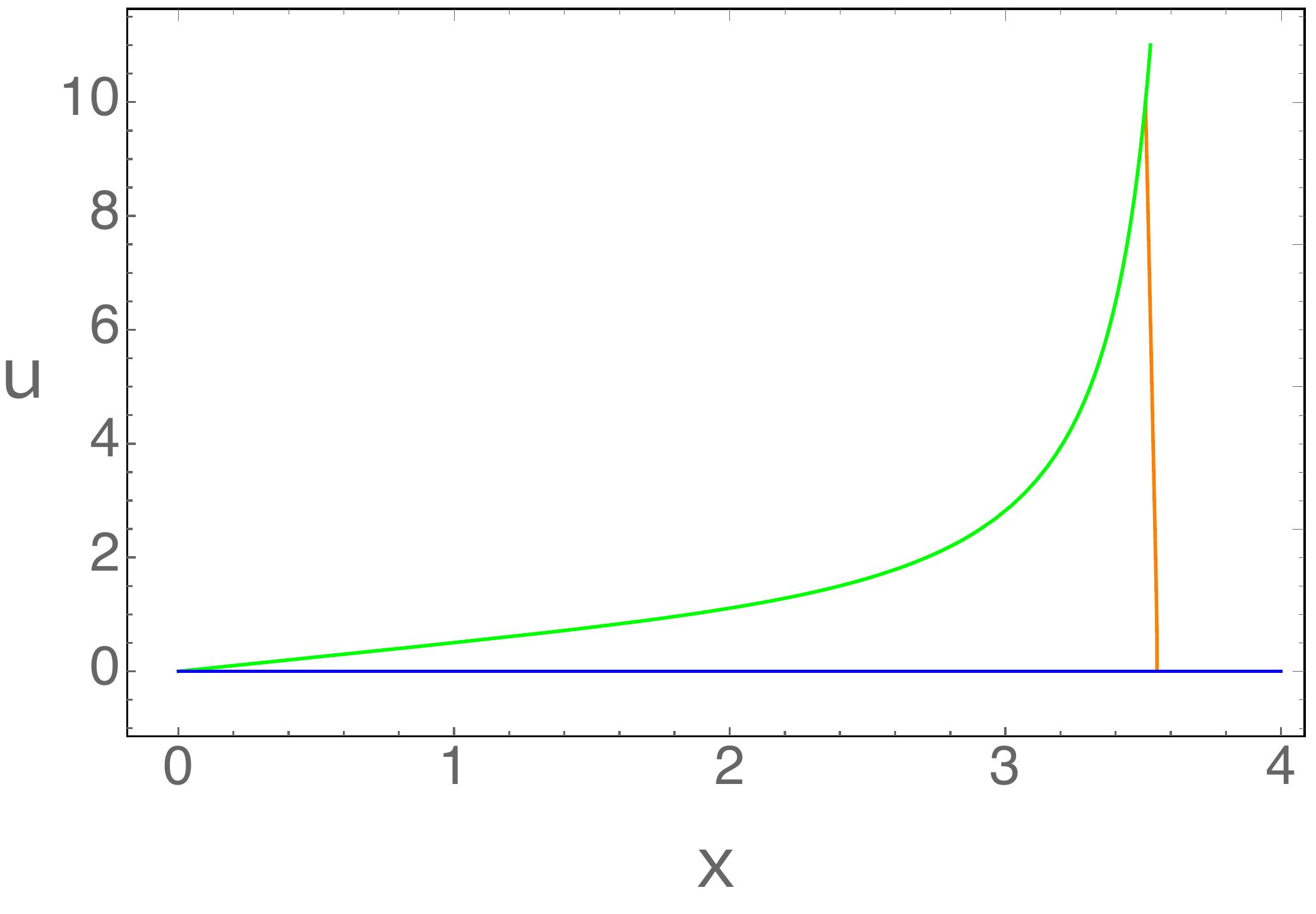}
\end{center}
\vspace{-0.3cm}
\caption{\small Three typical extremal surfaces for $c>0$: We have set $n=2, c=1$ ({\em left}), $n=4, c=1/2$ ({\em middle}),  and $n=4,c=2$ ({\em right}). The green line is the location of the EOW brane $Q$, and the blue line is the $x$ axis of BCFT. 
}
\label{fig:ee-sol}
\end{figure}

Three typical extremal surfaces are shown in Fig. \ref{fig:ee-sol}. The blue line is the $x$ axis of the BCFT, while the green line is the location of the EOW brane $Q$. The left plot is for $n=2$, and there always exists a single curved extremal surface for arbitrary $\ell$, shown as the orange line. The middle and right plots are for $n=4$ while different $c$. In the middle plot, $c<c_m$ and $x_m<\ell<\ell_m$, we have three extremal surfaces. In the right plot, $c>c_m$ while $\ell<x_m$, we have only one curved extremal surface. In the following, we will study the behavior of the entanglement entropy and the renormalized entanglement entropy for these typical behaviors. 

\begin{figure}[h!]
\begin{center}
\includegraphics[width=0.3\textwidth]{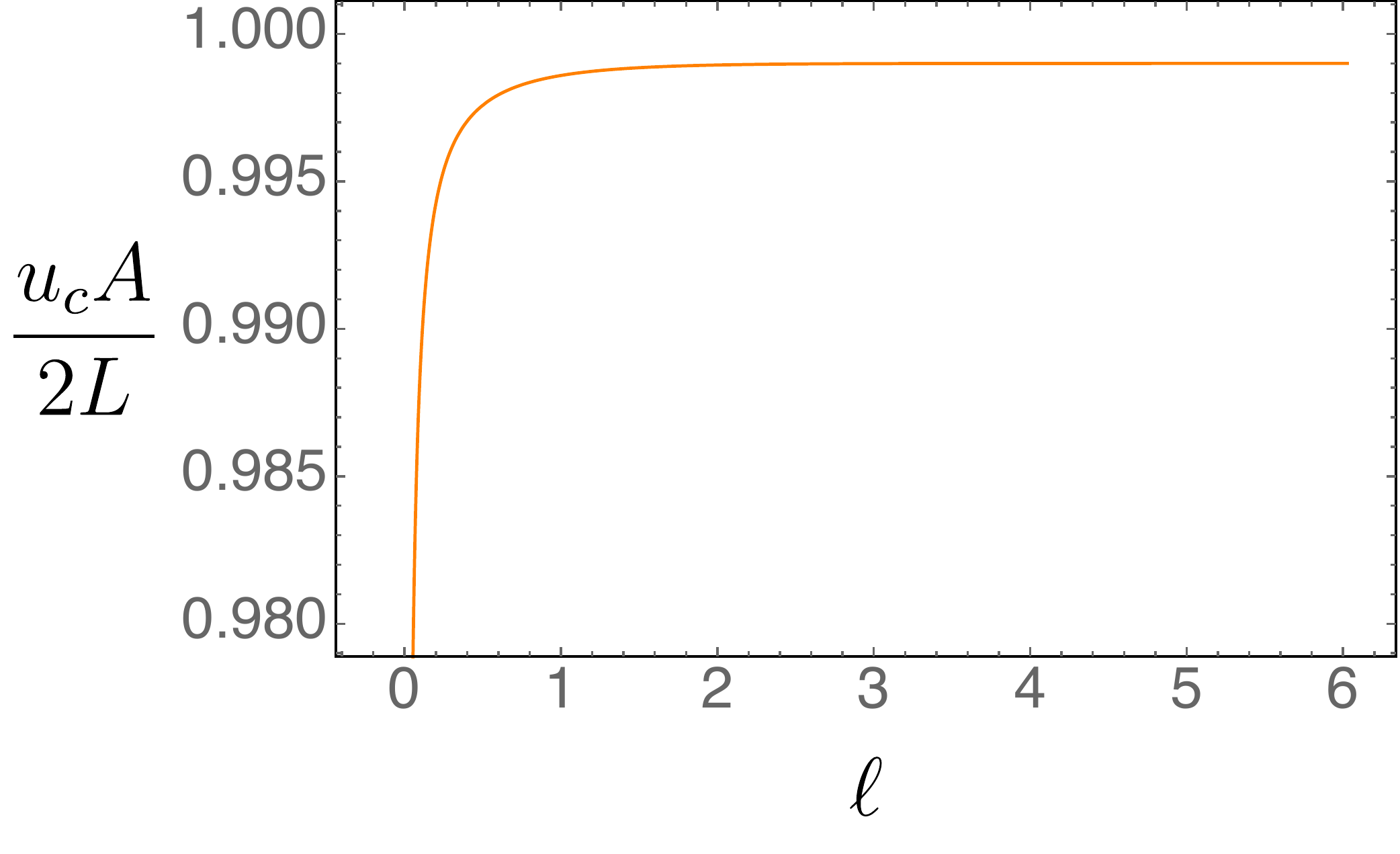}
\includegraphics[width=0.3\textwidth]{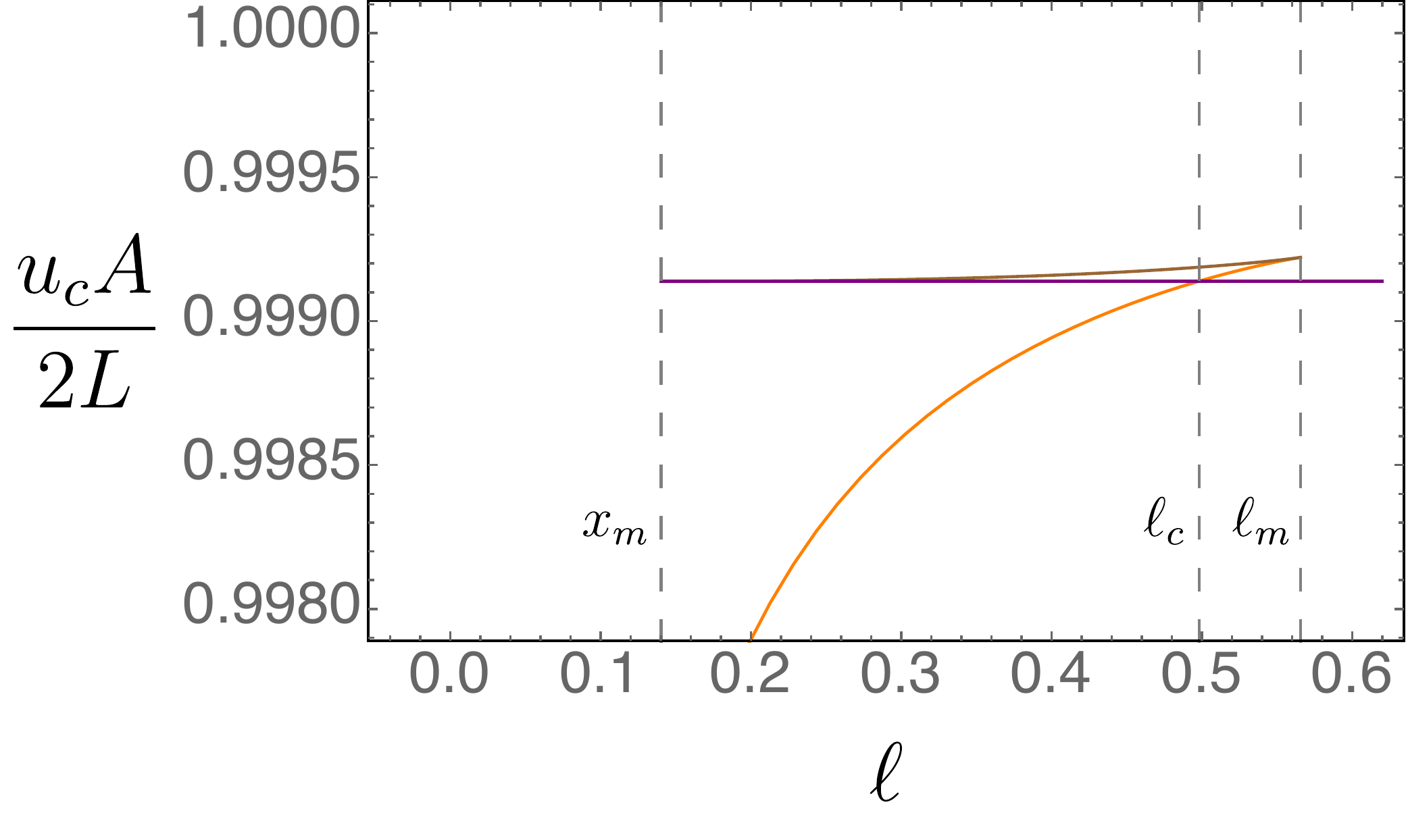}
\includegraphics[width=0.3\textwidth]{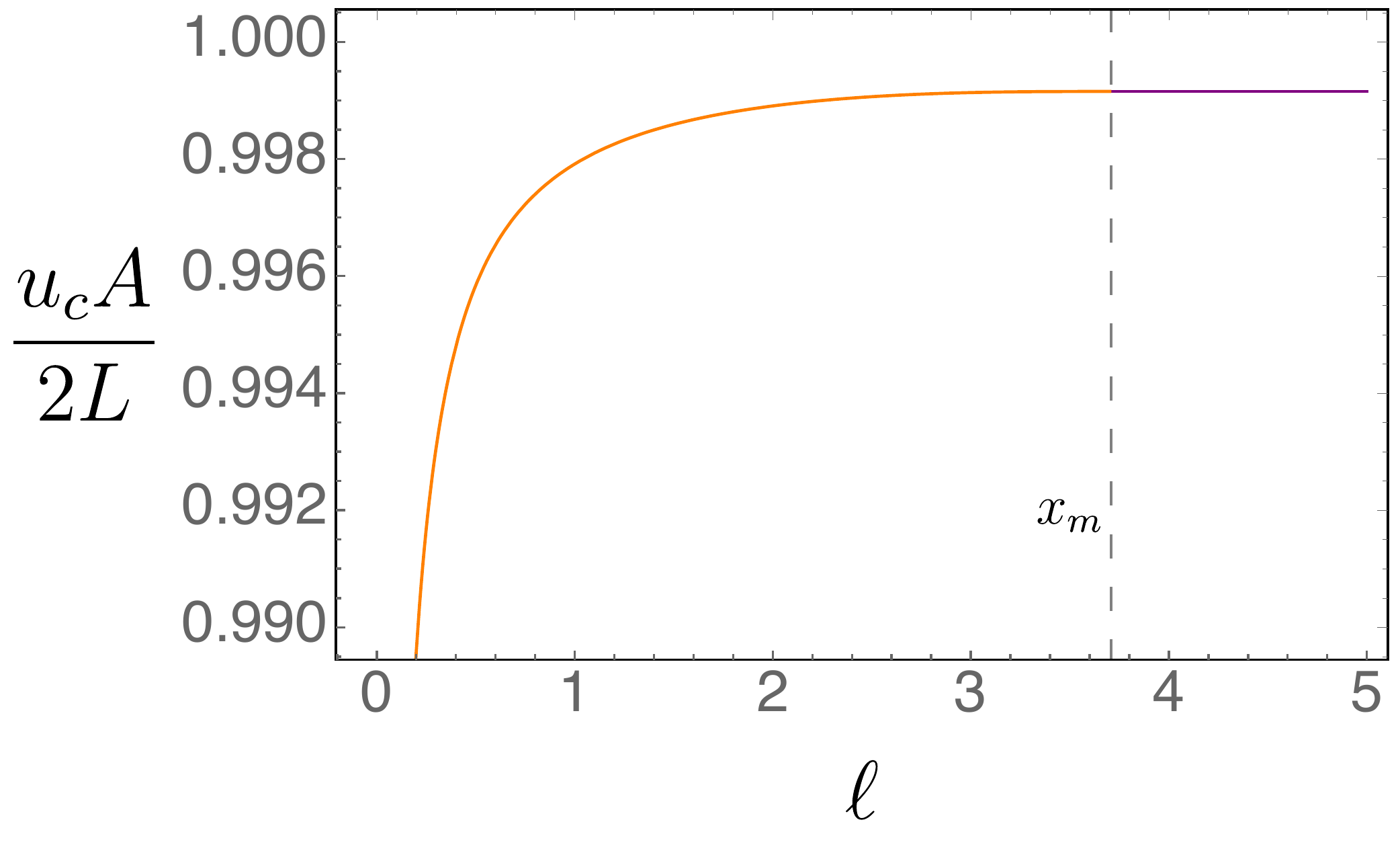}
\end{center}
\vspace{-0.4cm}
\caption{\small Plots of the area of extremal surfaces as a function of the width of the strip $\ell$ for $n=2, c=1$ ({\em left}), $n=6, c=1/10$ ({\em middle}), and $n=4,c=2$ ({\em right}). 
We have set $u_c=10^{-3}$. The one with the smallest area gives the correct entanglement entropy. 
}
\label{fig:ee-sol2}
\end{figure}

The entanglement entropy of the strip can be obtained from the area of the minimal surface 
\bea
\label{eq:ee2}
\begin{split}
S=\frac{A}{4G}&=\frac{L}{2G}\,\int_{x_*}^{\ell-\epsilon}\, dx\frac{1}{u^2}\,\sqrt{1+\frac{u'^2}{f}}\,.
\end{split}
\eea
When there are multiple extremal surfaces, we again need to choose the one with the minimal area. 
The areas of the above typical configurations of extremal surfaces can be found in Fig. \ref{fig:ee-sol2}. For $n=2$ and $c=1$ (left), the entanglement entropy is continuous and smooth when we increase $\ell$. For $n>2$, we find that, for the case $c<c_m$ (middle) or $c>c_m$ (right), there is a continuous transition at $\ell_c$ (with $x_m<\ell_c<\ell_m$) or smooth crossover  at $\ell=x_m$ from the orange curved line to the purple straight line. 

\begin{figure}[h!]
\begin{center}
\includegraphics[width=0.3\textwidth]{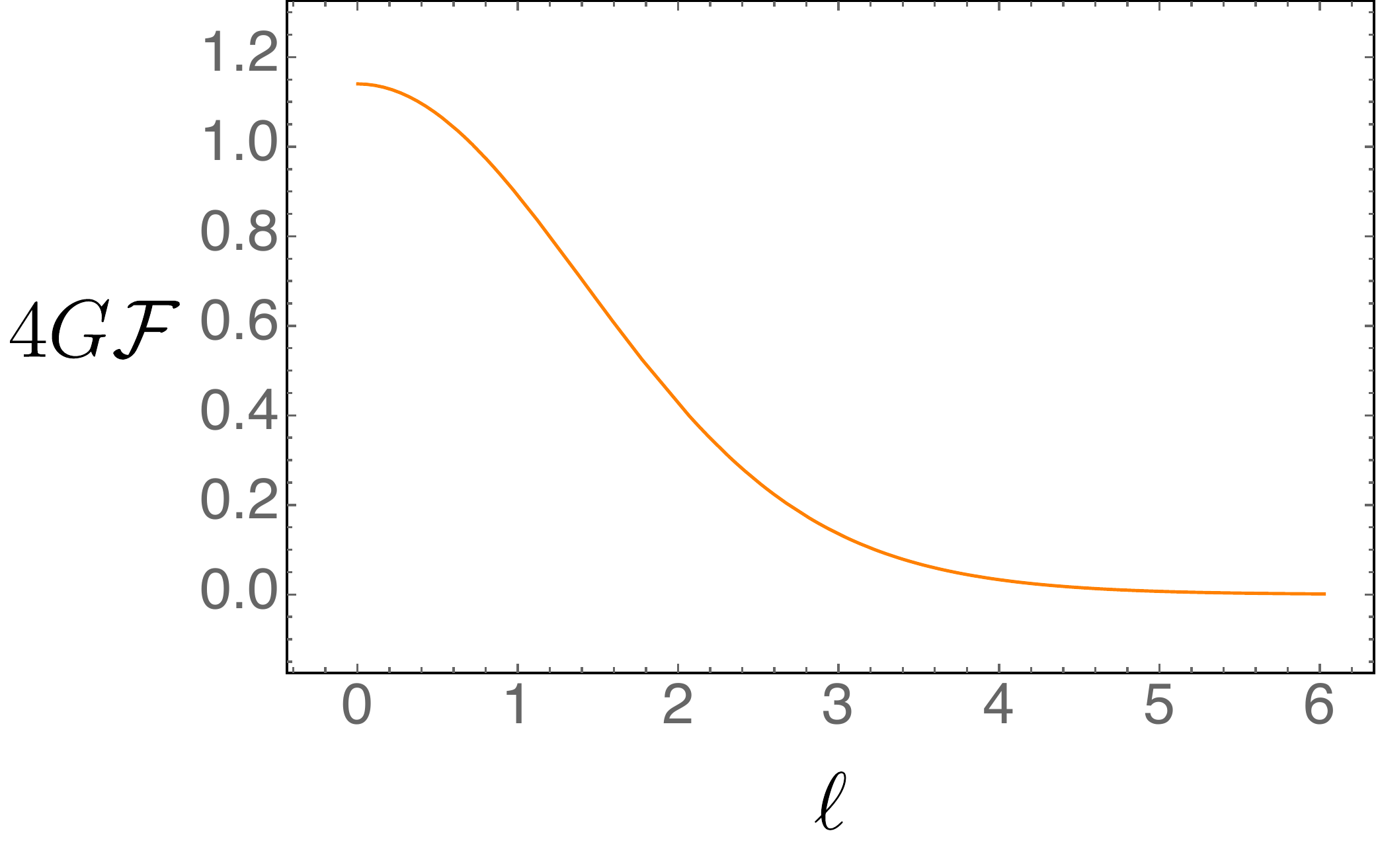}
\includegraphics[width=0.3\textwidth]{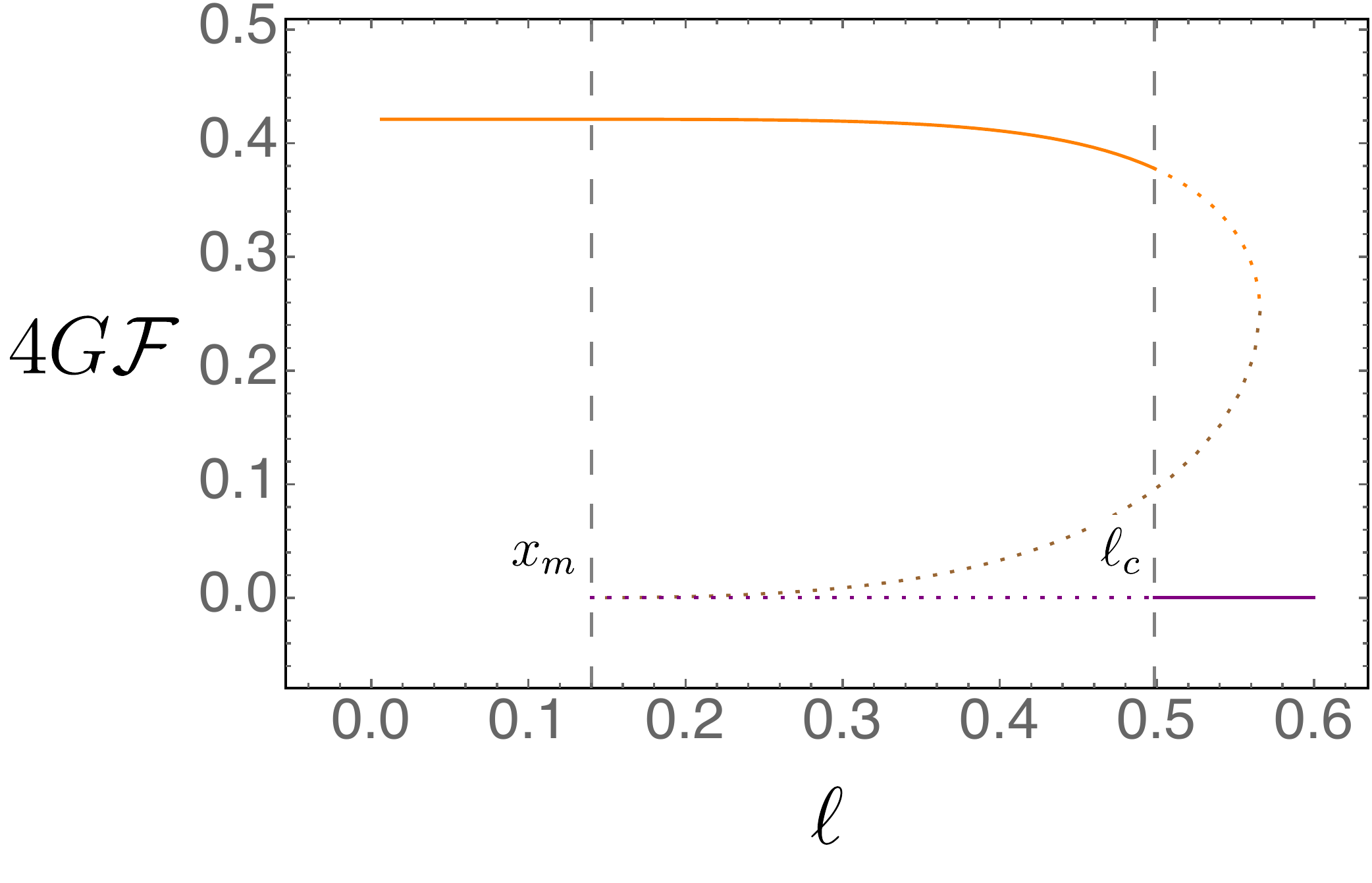}
\includegraphics[width=0.3\textwidth]{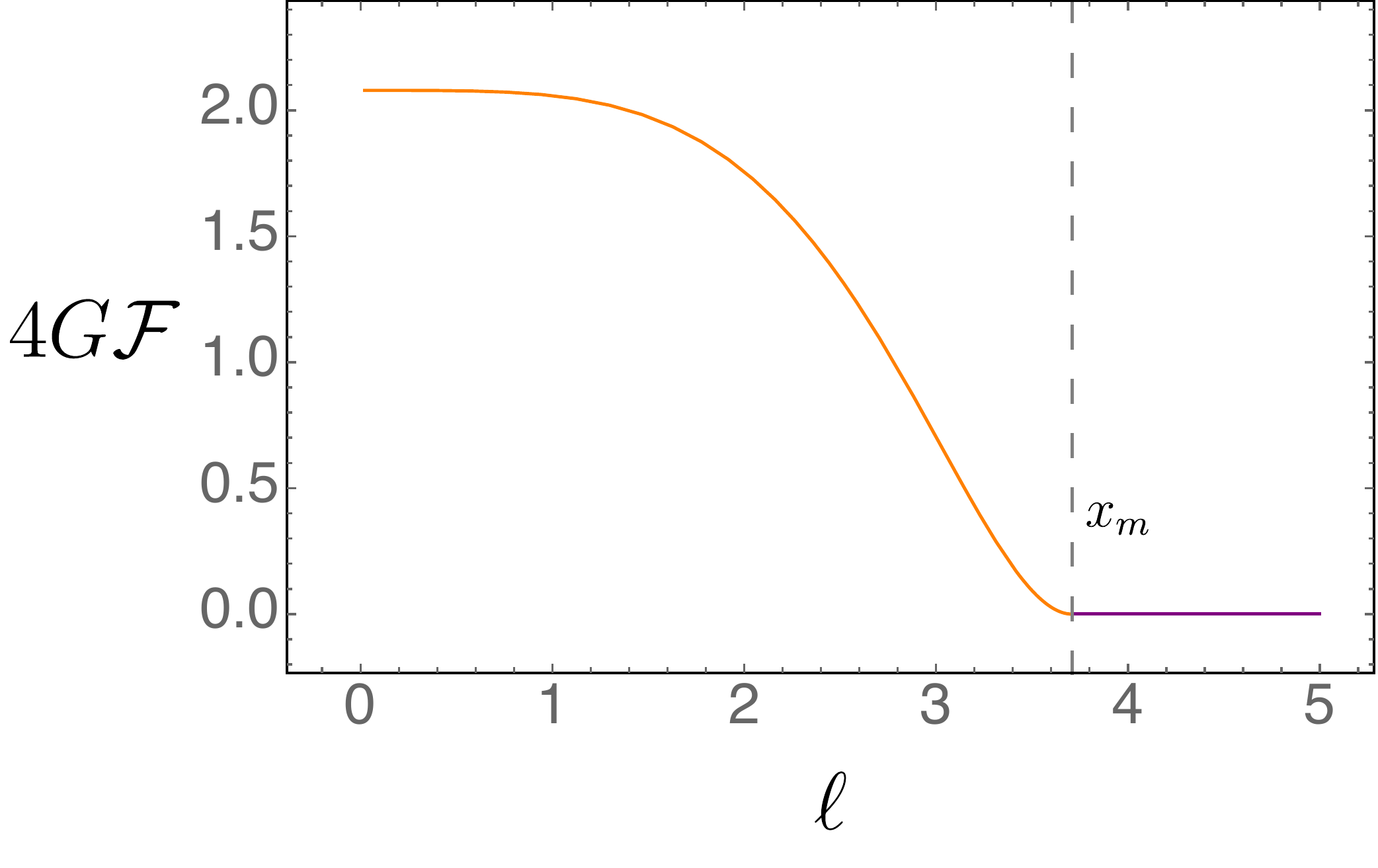}
\end{center}
\vspace{-0.4cm}
\caption{\small Plots of the renormalized entanglement entropy $4G\mathcal F$ as a function of $\ell$ for $n=2, c=1$ ({\em left}), $n=6, c=1/10$ ({\em middle}), and $n=4,c=2$ ({\em right}). 
}
\label{fig:ree-sol}
\end{figure}

Because of the divergence of the entanglement entropy, we again study the renormalized entanglement entropy which is independent of the cutoff. Close to $u\to 0$, from \eqref{eq:eerel3} we have 
\be\label{eq:uto02}
x(u)=\ell-\frac{u^3}{3 u_*^2\,(1+c^2)}+\cdots\,.
\ee
Following \cite{Myers:2012ed,Liu:2013una} and the calculations in appendix \ref{app:ree}, we do the variation of \eqref{eq:ee2} with respect to $\ell$. Using \eqref{eq:uto02}, we can obtain  the dimensionless renormalized entanglement entropy 
\bea 
\label{eq:effent2}
\mathcal{F}=
 \frac{\ell^2}{2L}\frac{\partial S}{\partial \ell }=\frac{1}{4G}\frac{\ell^2}{u_*^2\sqrt{1+c^2}}\,.
\eea
The behavior of renormalized entanglement entropy is shown in Fig. \ref{fig:ree-sol}. The left plot is for $n=2$, and we find that the renormalized entanglement entropy is non-negative and  monotonically decreasing. 
The middle plot is for $n=6$ while $c<c_m$. The solid line is for the configuration with minimal area, and we see that there is a jump for the renormalized entanglement entropy at $\ell=\ell_c$ (with $x_m<\ell_c<\ell_m$). The right plot is for $n=4$ while $c>c_m$. We see that the renormalized entanglement entropy is continuous while not smooth at $\ell=x_m$ (black dashed line).

\begin{figure}[h!]
\begin{center}
\includegraphics[width=0.6\textwidth]{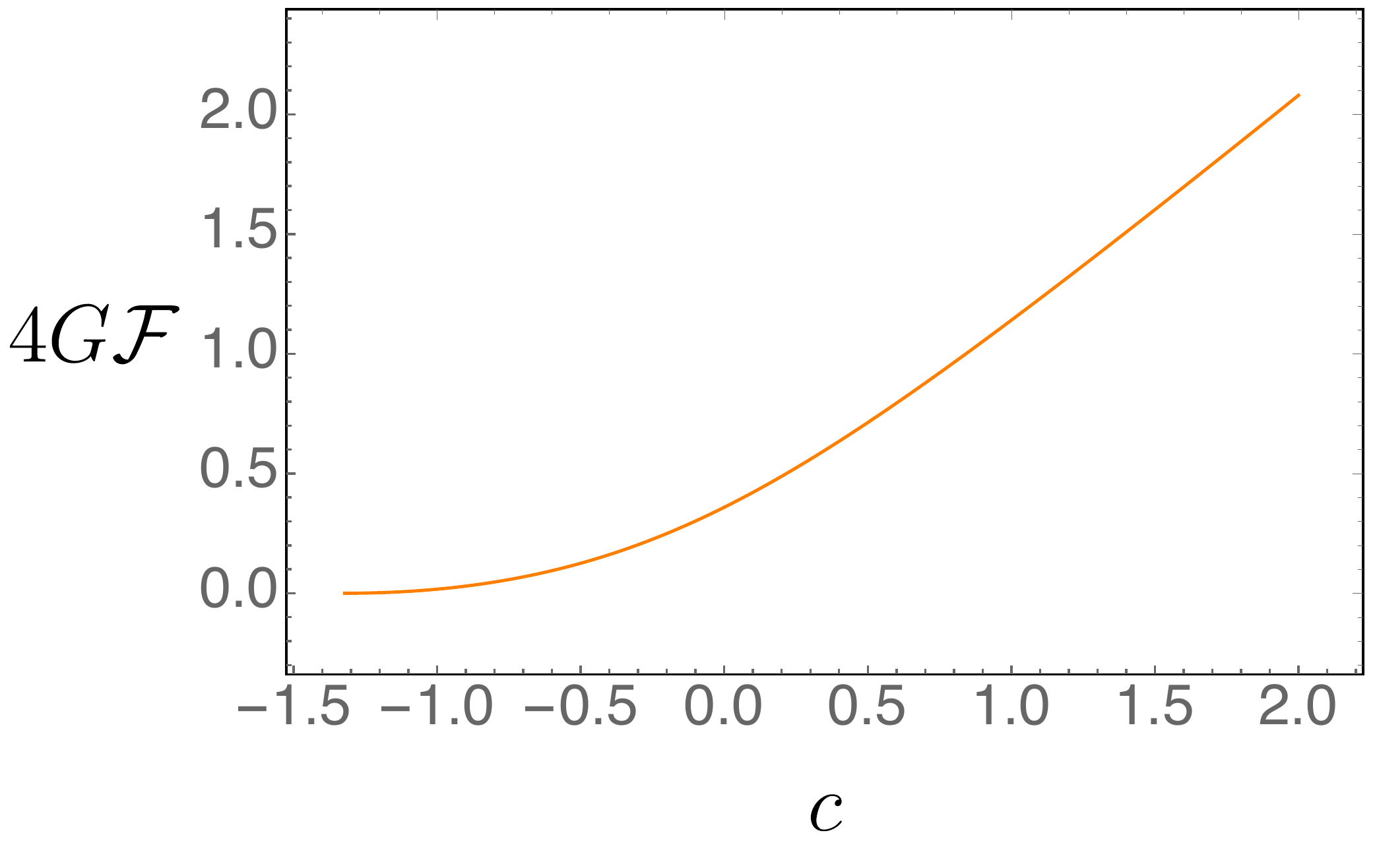}
\end{center}
\vspace{-0.5cm}
\caption{\small Plots of $4G\mathcal F(\ell\to 0)$ as a function of $c$. The behavior is independent of $n$, since in the limit $\ell\to 0$ only the UV physics is important.
}
\label{fig:ree-c}
\end{figure}

Finally, let us discuss the effect of the boundary on the BCFT. We have seen that, with different choices of $c$ which is related to the effective tension of the EOW brane, the profiles of the brane are different. From the plot in Fig. \ref{fig:reeF}, the right one in Fig. \ref{fig:ee0n2}, and the left one in Fig. \ref{fig:ree-sol}, which are for the same bulk geometry with $n=2$ while different values of $c$, we find that the renormalized entanglement entropy behaves differently. Fig. \ref{fig:ree-c} shows the behavior of the renormalized entanglement entropy in the limit $\ell\to 0$ as a function of $c$ which is independent of $n$. It is known that the renormalized entanglement entropy can be viewed as the number of the effective degrees of freedom. We find that, with different profiles of the EOW brane which should determined by the properties of the BCFT, the UV degrees of freedom of the BCFT are different. When the size of the strip is large enough, i.e., $\ell>\ell_c$, the renormalized entanglement entropy goes to zero. 

\section{AdS Soliton in AdS/BCFT}
\label{sec3}

In the previous section, we have studied the gapped system which is dual to Einstein-scalar theory in the presence of a boundary. In this section, we will study another gapped system which is described by the AdS soliton \cite{Witten:1998zw}. 

We will focus on the five-dimensional AdS soliton where one of the spatial dimensions is compact; therefore, the dual field theory is 2+1 dimensional with an additional compact extra dimension. We consider the presence of a boundary for the dual field theory along the noncompact dimension and study its transports and entanglement structure parallel to the discussion in the previous section.   

The action of the holographic model is 
\bea
\mathcal{S}_\text{bulk}&=&\mathcal{S}_N+\mathcal{S}_Q\,,
\eea
where
\bea
\begin{split}
\mathcal{S}_N&=\int_N d^5x\sqrt{-g}\,
  \bigg[\frac{1}{2\kappa^2}\bigg(R+12 \bigg) -\frac{1}{4e^2}F^2  \bigg] \,,\\
\mathcal{S}_{Q}&=\int_Q d^4x\sqrt{-\gamma}\, \bigg[\frac{1}{\kappa^2}\big(K-T\big)
   \bigg]\,.
\end{split}
\eea
The EOW brane $Q$ is similar to the setup in the previous subsection as shown in Fig.  \ref{fig:cf}, which extends from the boundary $P$ of the BCFT to the bulk. $T$ is the tension of the brane. We set $2\kappa^2=e=1$. 

The equations of motion in $N$ are
\bea
R_{ab}-\frac{1}{2} g_{ab}\big(R+12\big)-\frac{1}{2} \bigg[\mathcal{F}_{ac}\mathcal{F}_{b}^{~c}-\frac{1}{4}g_{ab}\mathcal{F}^2\bigg]&=0 \,,\\
\nabla_b F^{ba}&=0\,.
\eea

The metric of AdS soliton 
geometry at zero density is 
\be
\label{eq:adssoliton}
ds^2=\frac{1}{u^2}\bigg[-dt^2+dx^2+dy^2+\frac{du^2}{f(u)}\bigg]+\frac{f(u)}{u^2}d\theta^2\,,~~ f(u)=1-\frac{u^4}{u_0^4}\,,~~~A_a=0\,.
\ee
$\theta$ has a period of $\theta\sim \theta +\pi u_0$. Note that $u_0$ sets the scale of the gap. This geometry is asymptotic to AdS and approaches to $R^{1,2}\times S^1$ near the boundary. Here, $M$ is also defined on the half plane with $x\geq 0$.
The AdS boundary is at $u\to 0$. AdS soliton exists at $u\leq u_0$.
Obviously, the AdS soliton geometry \eqref{eq:adssoliton} is a solution of the system. 

 The equations of motion on $Q$ are
\bea\label{eq:adssoliton-q}
\begin{split}
K_{\mu\nu}-(K-T)\gamma_{\mu\nu}&=0\,, \\
n_aF^{ab}&=0\,,
\end{split}
\eea
where $n^a$ is the outforward unit vector for $Q$. We assume $Q$ is described by equation $u=u(x)$, and then we have 
\bea
\label{eq:sol-na}
(n^t,n^x,n^y,n^u,n^\theta)=\bigg(0,~~\frac{-u}{\sqrt{1+f(u)x'(u)^2}}\,,~~0,~~\frac{u f(u)x'(u)}{\sqrt{1+f(u)x'(u)^2}}\,,~~0\bigg)\,.
\eea

Plugging \eqref{eq:sol-na} into the equations on $Q$, we find that there is only one consistent solution with trivial embedding $x(u)=0$ with $T=0$.\footnote{It is interesting to study if other nontrivial consistent embedding could be found when we choose Dirichlet or mixed boundary conditions. We leave this possibility for future study.}
This fact makes the discussion for the  AdS soliton simpler than the gapped geometry in section \ref{subsec:groundstate} where there are different profiles for $Q$. Note that here we assume that the geometry in $N$ is the AdS soliton and the brane $Q$ does not backreact the geometry, which results in a tensionless solution. It would be interesting to study the case of the  deformed AdS soliton solution by a finite tension brane, and we will not consider it here. Note that there exists the AdS soliton solution with the EOW brane of finite tension \cite{Fujita:2011}, where the
boundary of BCFT is defined along the compact spatial direction. However, in our case, the boundary of BCFT is defined along a noncompact spatial direction, in order to make
comparison with the discussion in the previous section.

\subsection{Conductivity}
\label{ss:solcon}

With the above configurations for the  AdS soliton with a boundary, we can study its transport physics and entanglement structure. We first study the conductivity along the $y$ direction. 
Considering the fluctuations of the gauge fields as \eqref{eq:gauflu}, we obtain 
the fluctuation equation for $a_y$ in $N$:  
\bea\label{eq:fulay2}
a_y''+\left(\frac{f'}{f}-\frac{1}{u}\right)a_y'+\frac{\omega^2+\partial_x^2}{f} a_y&&=0\,, 
\eea
and the equation for $a_y$ on the EOW brane $Q$:  
\be
(-\partial_x a_y +f x' \partial_u a_y)\Big{|}_Q=0\,.
\ee
Since $Q$ is described by $x=0$, the boundary equation can be simplified further as $\partial_x a_y\big{|}_Q=0$. 
Therefore, this is quite similar to the case of $c=0$ in section \ref{subsec:con}, and we have the solution 
\be a_y=c_0 a(u,\omega)\,,
\ee
where $c_0$ is a constant and $a(u,\omega)$ satisfy 
\bea \label{eq:fula}
a''+\left(\frac{f'}{f}-\frac{1}{u}\right)a'+\frac{\omega^2}{f} a=0\,. 
\eea

We have analyzed in appendix \ref{app:sch} that the system is gapped by transforming the above equations into a Schr\"{o}dinger problem to show that the real part of conductivity is a sum of discrete poles. Thus, for this model, in both $M$ and $P$ the conductivities along the boundary of BCFT are trivial.

\subsection{Entanglement entropy}

Similar to the discussions in section \ref{subsec:2ee}, we study the entanglement entropy of a strip geometry. The subsystem under consideration is $0<x<\ell$, while $-L<y<L$ with $L\to\infty$ and $0\leq \theta\leq \pi u_0 $. 

The extremal surface $\gamma$ is specified by $u=u(x)$, which is a section at $y=\text{const}$.
The induced metric on $\gamma$ is
\be
ds^2_\gamma=\frac{1}{u^2}\bigg[\,\Big(1+\frac{u'^2}{f(u)}\Big)\,dx^2+dy^2\bigg]+\frac{f(u)}{u^2}d\theta^2\,,
\ee
from which one obtains the area functional 
\be
\label{eq:areafun2}
A=2\pi u_0 L\,\int_{x_*}^{\ell} dx\,\frac{\sqrt{f+u'^2}}{u^3}\,.
\ee 

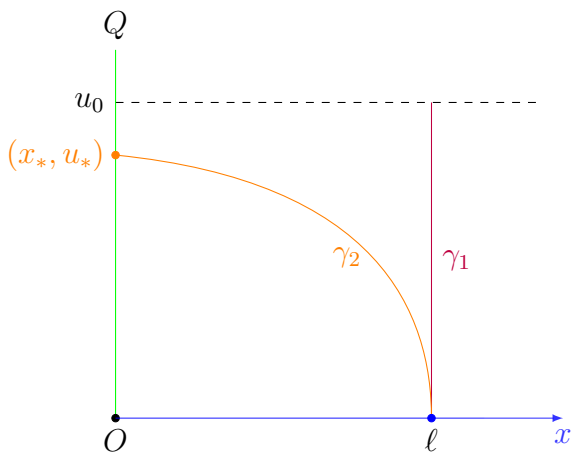
\begin{figure}[h]
\begin{center}
\begin{tikzpicture}[scale=0.7]
\draw[black] (0,0) node[anchor=north, below]{ $O$ };
\draw [blue!10] (0,0)--(7,0);
\draw [black, dashed] (0,6)--(8,6);
\draw [purple] (6,0)--(6,6) node[midway, right]{$\gamma_1$};
\draw [green] (0,0)--(0,7); 

\draw [orange] (6,0) .. controls (6,2) and (5,4.5) .. (0,5) node[left]{$(x_*, u_*)$ } node[midway, left]{$\gamma_2$};

\draw[black] (0, 7)  node[above]{ $Q$ };
\draw[black] (0, 6)  node[left]{ $u_0$ };
\filldraw[orange] (0, 5) circle (2pt);
\filldraw[black] (0, 0) circle (2pt);
\filldraw[blue] (6, 0) circle (2pt);
\draw[black] (6, 0)  node[below]{ $\ell$ };

\draw[-latex, blue, opacity=0.8] (0, 0)--(8.5, 0) node[anchor=north, at end]{ $x$ };

\end{tikzpicture}
\end{center}
\vspace{-0.3cm}
\caption{\small Cartoon plot for the extremal surfaces $\gamma_1$ and $\gamma_2$ in the AdS soliton geometry with an EOW brane $Q$. }
\label{fig:caree-sol}
\end{figure}

Since the above functional does not implicitly depend on $x$, there is a conserved quantity 
\be
\label{eq:consq2sol}
\frac{f}{u^3\sqrt{f+u'^2}}=C\,.
\ee
Note that, for the extremal surface, we have boundary $u(\ell)=0$. Since the geometry is only for $0<u\leq u_0$ and $Q$ is located at $x=0$, one might expect that there are two different kinds of extremal surfaces as shown in the cartoon plot in Fig. \ref{fig:caree-sol}. 

The first configuration $\gamma_1$ in Fig. \ref{fig:caree-sol} is the surface $x=\ell$, which corresponds to $C=0$ in \eqref{eq:consq2sol}. Note that this configuration should exist for any $\ell$. The entanglement entropy from this extremal surface is 
\begin{align}
\label{eq:ee-sol-straight}
S=\frac{2\pi u_0 L}{4G}\int_{u_c}^{u_0} \frac{du}{u^3}\,
=\frac{\pi u_0 L}{4G}\left(\frac{1}{u_c^2}-\frac{1}{u_0^2}\right) ,
\end{align}
where $u_c$ is the cutoff close to the boundary and $u(\ell-\epsilon)=u_c$. 
When $u_c\to 0$, we have $u_c A/(2\pi u_0 L)\to 1/2$. 
The entanglement entropy is independent of $\ell$, and, therefore, we have 
$\partial{S}/\partial \ell=0$.

\begin{figure}[h!]
\begin{center}
\includegraphics[width=0.44\textwidth]{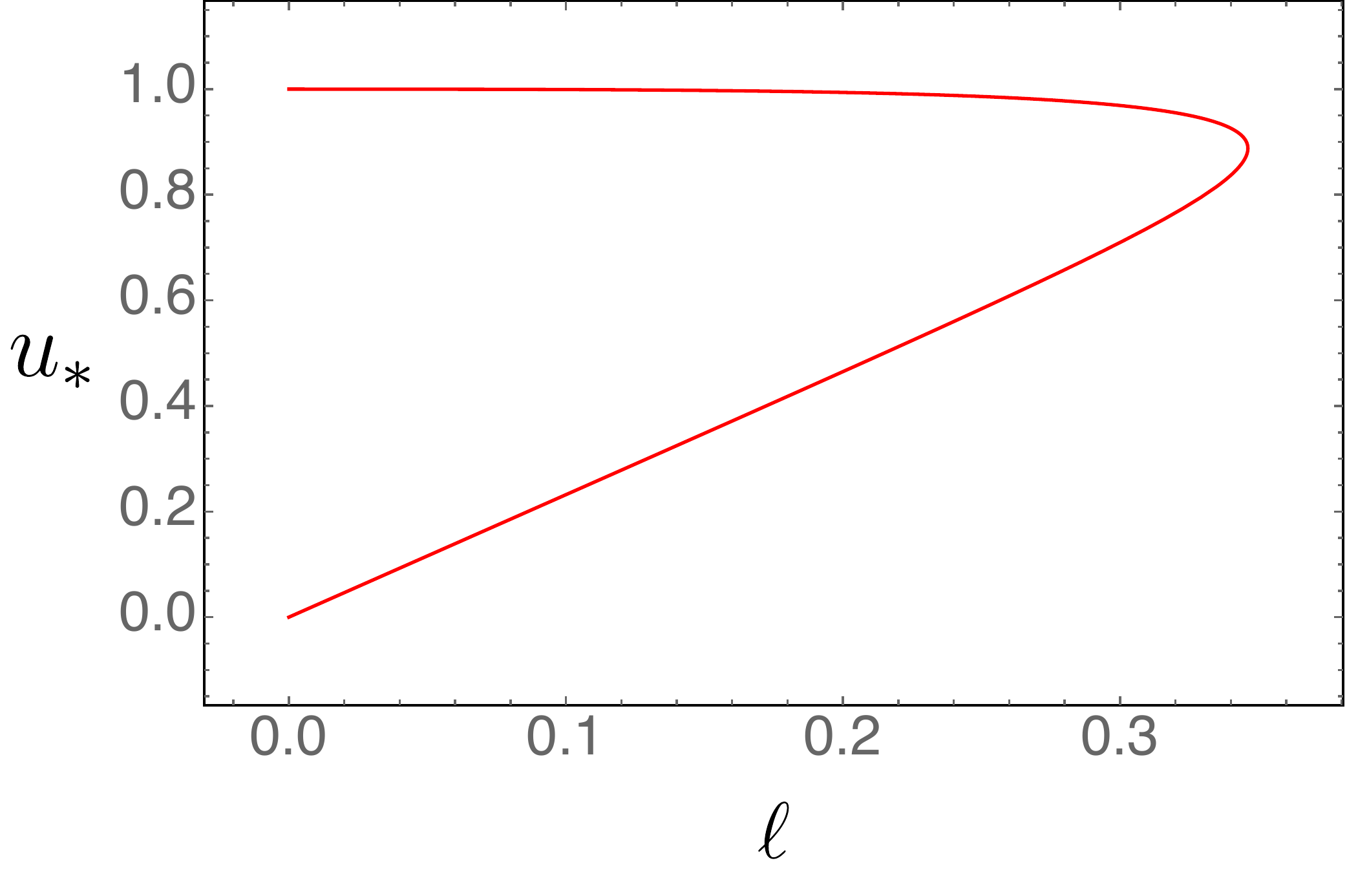}
~~~
\includegraphics[width=0.43\textwidth]{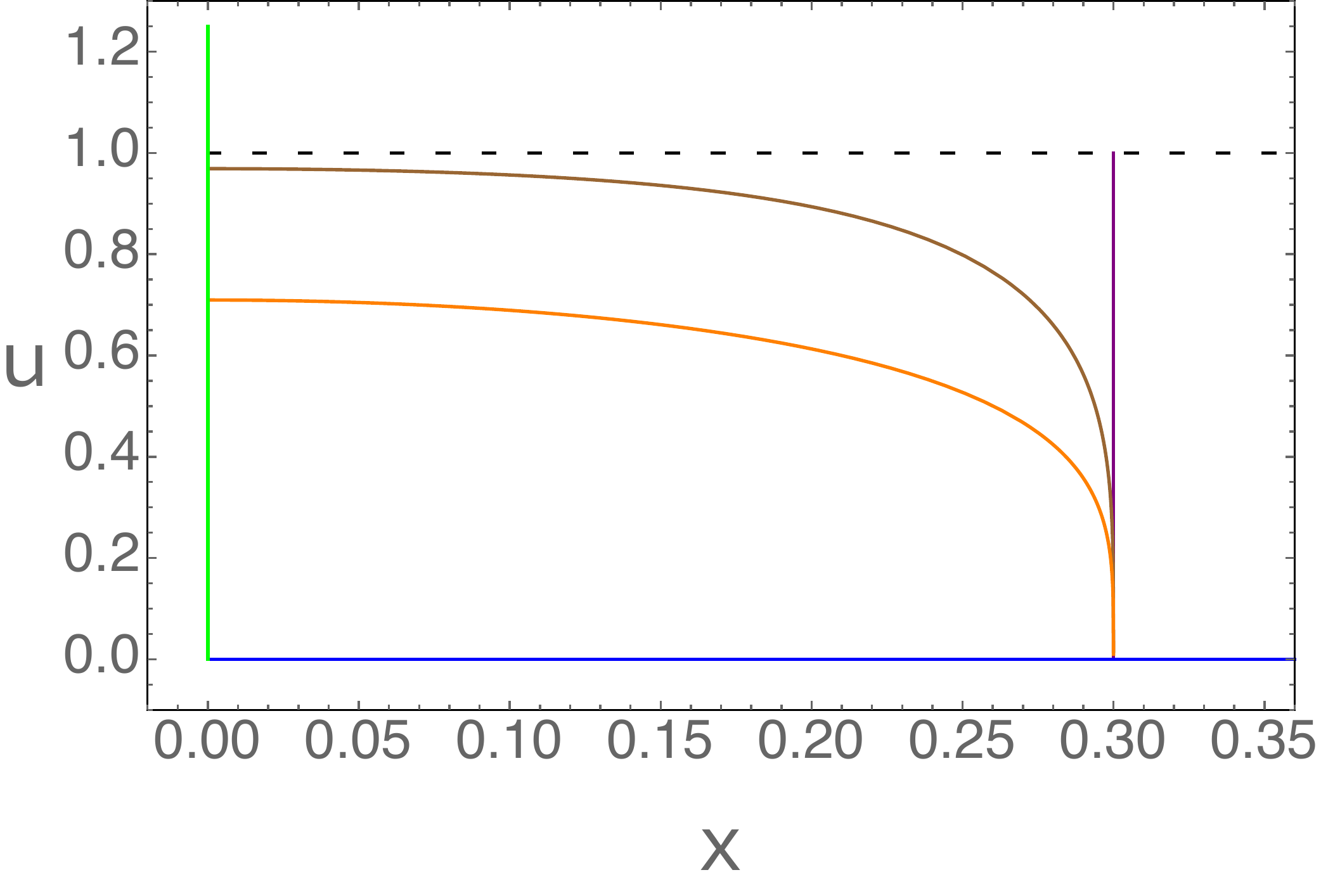}
\end{center}
\vspace{-0.3cm}
\caption{\small  {\em Left:} the location of intersecting point $u_*$ as function of the width of the strip $\ell$. {\em Right:} one example of the extremal surfaces in AdS soliton geometry with the EOW brane for a given width $\ell<\ell_m$. Note that here we have set $u_0=1$.
}
\label{fig:config-s1}
\end{figure}

Another configuration $\gamma_2$ in Fig. \ref{fig:caree-sol} exists only for small $\ell$. We have the intersecting point $(x_*, u_*)$ between the extremal surface and the EOW brane where $n_Q\cdot n_\gamma=0$, i.e.,
$u'(x_*=0)=0$.  
From \eqref{eq:consq2sol}, we have $C=\sqrt{f(u_*)}/u_*^3$, which leads to 
\be
\label{eq:eerel2sol}
u'=-\sqrt{\frac{u_*^6 f^2}{u^6 f(u_*)}-f}\,.
\ee
Then we have the relation 
\be
\label{eq:eeeqn1sol}
\ell=\int^{u_*}_0 du\, \frac{1}{\sqrt{\frac{u_*^6 f^2}{u^6 f(u_*)}-f}}\,.
\ee
From \eqref{eq:eeeqn1sol}, one could obtain $u_*$ as a function of $\ell$ as 
shown in the left plot in Fig. \ref{fig:config-s1}. 
Note that we should have $u_*< u_0$. There exists a maximal value $\ell_m$ below which we have two different configurations of curved extremal surfaces. One example of the extremal surfaces at $\ell<\ell_m $ is shown in the right plot in Fig. \ref{fig:config-s1}. These features remind us of the discovery in \cite{Klebanov:2007ws}  without a boundary (see also, e.g., \cite{Jokela:2020wgs}). 

\begin{figure}[h!]
\begin{center}
\includegraphics[width=0.45\textwidth]{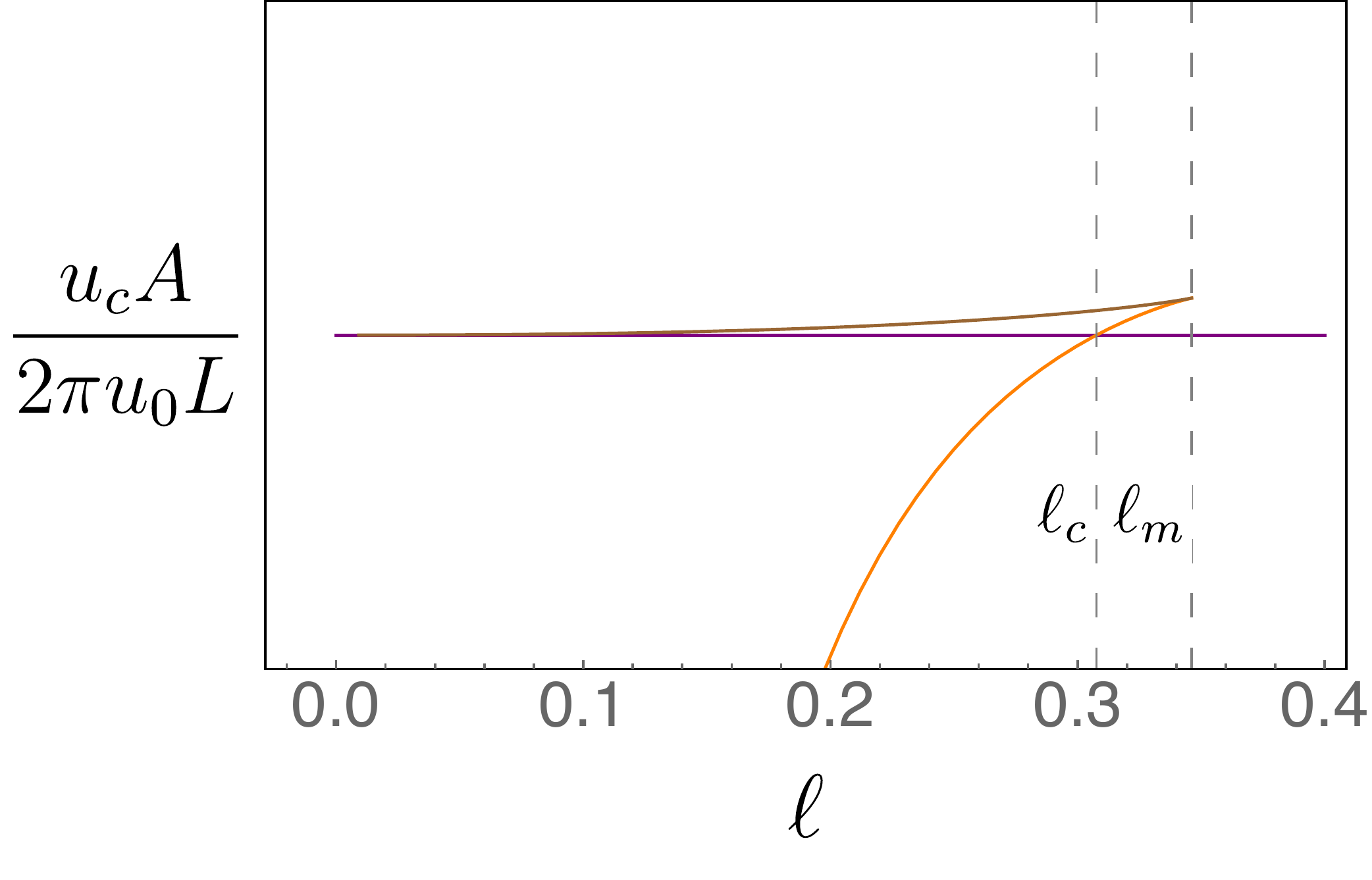}
~~
\includegraphics[width=0.48\textwidth]{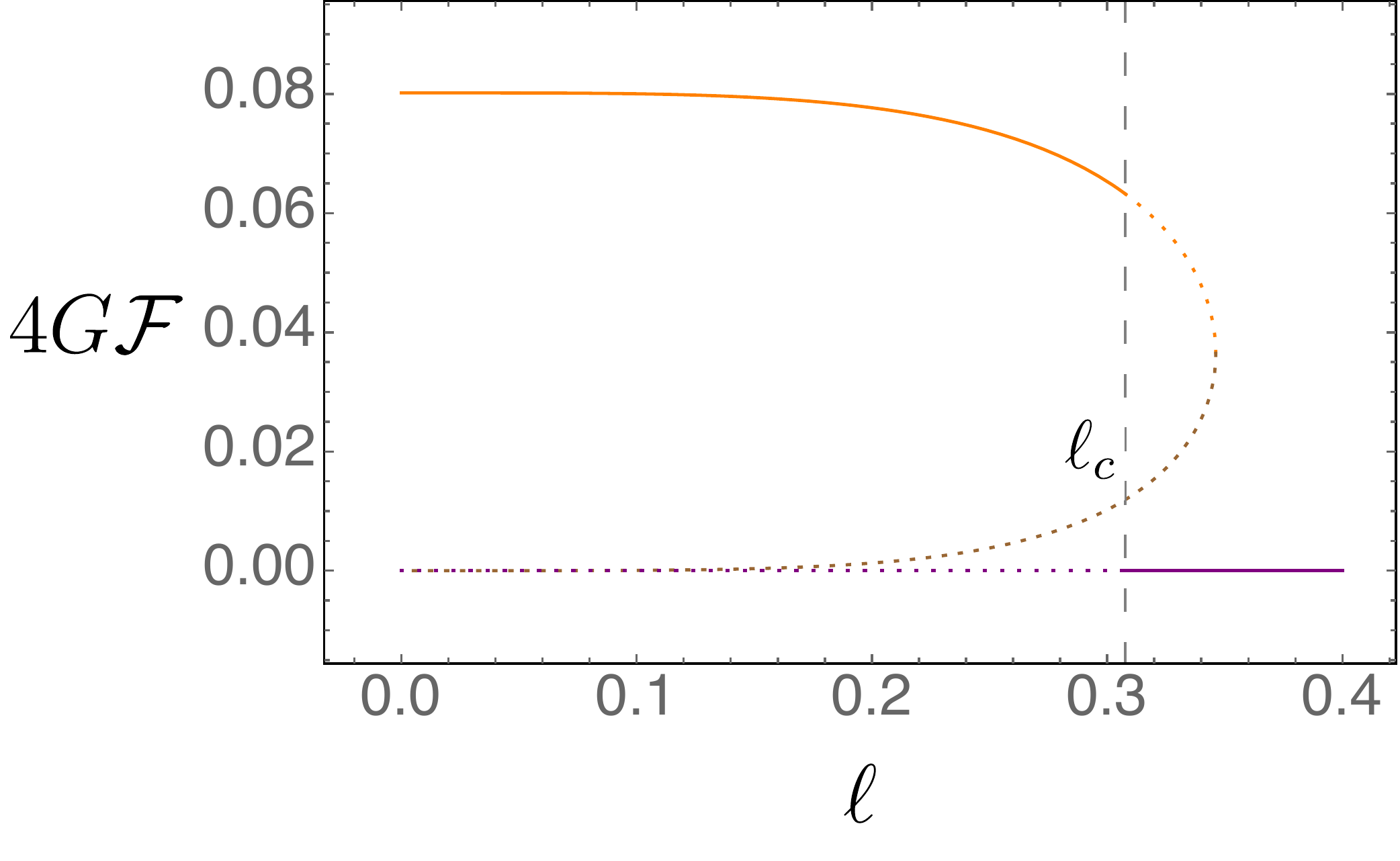}
\end{center}
\vspace{-0.3cm}
\caption{\small  {\em Left:} the area of the extremal surfaces $u_cA/(2\pi u_0 L)$ as function of the width of the strip $\ell$. When we set the cutoff $u_c=10^{-5}$, the difference along the vertical axis is of the order of $10^{-10}$, and, thus, we do not explicitly show them.  {\em Right:} the renormalized entanglement entropy $4G\mathcal{F}$ as a function of the width $\ell$. There is a discontinuous transition for renormalized entanglement entropy at $\ell=\ell_c$. 
}
\label{fig:ree-soliton}
\end{figure}

The entanglement entropy can be obtained from the area of the extremal surfaces 
\be
\begin{split}
\label{eq:eesol}
S&= \frac{A}{4G}=\frac{\pi u_0 L}{2G} \,\int_{x_*}^{\ell-\epsilon} dx\frac{\sqrt{f+u'^2}}{u^3}\,.
\end{split} 
\ee
The extremal surface with  minimal area gives the correct entanglement entropy. In the left plot in Fig. \ref{fig:ree-soliton}, we show the area of the extremal surfaces as a function of the width of the strip $\ell$. We find that there exists a critical value of $\ell_c$ which is smaller than $\ell_m$ that was found in the left plot in Fig. \ref{fig:config-s1}. When $\ell<\ell_c$,  the orange curved extremal surface has minimal area.  When $\ell>\ell_c$, the straight vertical purple curve has minimal area. Furthermore, when $\ell\to 0$, we have $f\to 1, u'\to-\infty$; from \eqref{eq:eesol}, we found that $S\propto \frac{1}{\ell^2}$, which reflects the UV properties of the BCFT. 

The renormalized entanglement entropy can also be discussed. Close to $u\to 0$, from \eqref{eq:eerel2sol} we have 
\be\label{eq:uto02sol}
x(u)=\ell-\frac{u^{4}}{4 u_*^3}+\cdots\,.
\ee
From the variation of \eqref{eq:eesol} with respect to $\ell$ and using \eqref{eq:uto02sol}, we obtain
\bea 
\label{eq:effentsol}
\mathcal{F}=
 \frac{\ell^3}{2\pi u_0 L}\frac{\partial S}{\partial \ell } =\frac{\sqrt{f(u_*)}}{4G}\frac{\ell^3}{u_*^3}\,.
\eea
The behavior of renormalized entanglement entropy is shown in the right plot in Fig. \ref{fig:ree-soliton}, which reminds us of the plot in Fig. \ref{fig:reeF}. We find that the renormalized entanglement entropy (solid lines) is non-negative and monotonically decreasing and there is a discontinuous transition at $\ell_c$ (dashed black line). Different from the discussion in the previous section, there are no free parameters similar to the effective tension of the EOW brane to make the entanglement structure more richer.  

\section{Conclusion and discussion}
\label{sec4}

We have studied the properties of two holographic gapped systems at zero density in the presence of boundaries using AdS/BCFT. The first gapped system is described by Einstein-scalar gravity, and the second one is the dual of AdS soliton geometry. In the first system,  the profiles of the EOW brane are quite  richer, depending on the effective tension of the EOW brane, while in the second system we find only one consistent trivial profile of the EOW brane. In these two systems, both the bulk and boundary conductivities in BCFT along the spatial direction of its boundary are trivial, and, hence, we learn that the strong correlation cannot make a topologically trivial insulator topologically nontrivial. The entanglement structure in the first system is very rich. The boundary physics has nontrivial effects on the entanglement structure of the system. For example, by comparing the plot in Fig. \ref{fig:reeF}, the right one in Fig. \ref{fig:ee0n2}, and the left one in Fig. \ref{fig:ree-sol}, which are for the same bulk geometry parametrized by $n=2$ while different values of $c$ that parametrized the effective tension of the EOW brane, we find that the renormalized entanglement entropies behave differently. Nevertheless, in the presence of a boundary, the renormalized entanglement entropy is always non-negative and monotonically decreasing and can discontinuously, continuously, or even smoothly evolve when we increase the length scale of the subsystem.
In the system of AdS soliton with a boundary, the renormalized entanglement entropy exhibits a ``unique" behavior with a discontinuous drop when we increase the size of the subsystem. 

One immediate open question is to study other fluctuations modes, e.g., metric fluctuations, or probing fermions, to check if there are any gapless degrees of freedom on the boundary of BCFT. 
Another interesting question is to define a proper physical quantity to extract the contribution from the boundary side. A possible candidate might be the quantity of the differences between the cases with and without boundaries. The entanglement structure in the gapped geometry without any EOW brane has been studied in \cite{Liu:2013una} where the renormalized entanglement entropy has been calculated to be the same expression as (\ref{eq:effent}). For the geometry without any boundary, one might naively identify the system through a mirror reflection $x\to -x$, which would result in the same conclusions as $c=0$. For $n=2$, we have seen that the renormalized entanglement entropy crucially depends on the value of $c$. Especially for small $\ell$ when  $c$ is positive, we have larger renormalized entanglement entropy, while when $c$ is negative, we have smaller renormalized entanglement entropy. This indicates that the different profiles of the EOW brane can add or reduce the UV degrees of freedom of the CFT.  It would be interesting to study how to reveal this procedure more precisely. Meanwhile, it would be interesting to study the entanglement entropy for other different subsystems to see if similar phenomena could be observed. 

It would be very interesting to construct a holographic model for a topological insulator from AdS/BCFT and then make predictions from the model. The study in this work suggests that new ingredients should be incorporated if we start from a gapped system by introducing an EOW brane using AdS/BCFT. One possibility is to study other types of gauge theories on the gravity side, e.g., the Dirac-Born-Infeld action for the gauge field, to make the boundary equation for the gauge field more complicated in order to have gapless excitations. Another possibility is to introduce a new dynamical gauge field on the brane to model the gapless excitations in analogy to the holographic Kondo model \cite{Erdmenger:2015xpq, Andrei:2018die}. 

Finally, it would be very interesting to understand the Island/BCFT correspondence \cite{Suzuki:2022xwv} for the gapped systems studied in this work.  We leave these interesting questions for  future research.  

\vspace{.3cm}
\subsection*{Acknowledgments}
 We thank Li Li, Rong-Xin Miao, Francisco Peña-Benitez, Jie Ren, Ya-Wen Sun, and Xin-Meng Wu for useful discussions. This work is supported by the National Natural Science Foundation of China Grant No.11875083. Jun-Kun Zhao is also supported by the National Natural Science Foundation of China Grant No.12122513 and No.12075298.

\vspace{.6 cm}
\appendix
\section{Analysis of conductivity in terms of Schrodinger equation}
\label{app:sch}

In this appendix, we solve the fluctuation equations for the cases of $c=0$ in gapped geometry and of AdS soliton by writing them into a Schr\"{o}dinger problem. 
The fluctuation equation can be written as
\be
\label{eq:appflu}
\bigg(\sqrt{f}Z a_y'\bigg)'+
\frac{Z}{\sqrt{f}} \omega^2 a_y=0\,. 
\ee
The above equation is the one we need to solve for $c=0$ in Sec. \ref{subsec:con}. 
Note that, by choosing $Z=\frac{ \sqrt{f} }{u}$, the above equation can also be reduced to (\ref{eq:fula}) for the case of the AdS soliton. 

Introducing $\xi$ and $v$ with 
\be 
\label{eq:newcoor}
\frac{d\xi}{du}=\frac{1}{\sqrt{f}}\,,~~~~ v=\sqrt{Z} a_y\,,\ee 
the above equations can be written into a Schr\"{o}dinger equation:
\bea
-\frac{d^2 v}{d\xi^2}+V_\text{eff} \, v=\omega^2 v
\eea
with effective potential
\bea
V_\text{eff}= \frac{f}{2Z}Z''-\frac{f}{4Z^{2}}Z'^2+\frac{f'Z'}{4Z}\,. 
\eea
with primes the derivatives with respect to $u$. 

\subsection{Gapped geometry in Sec. \ref{subsec:con}}

In the IR region, i.e., $u\to \infty$, from \eqref{eq:bgsol1} the background field behaves as  
\be f\sim u^n\,,~~~ \phi\sim \sqrt{2n}\log{u}\,.\ee 
From the expression \eqref{eq:conZ}, 
we have
\bea
V_\text{eff}=\frac{\alpha^2}{4}\,,
\eea
which is a constant from UV to IR. The Schr\"{o}dinger problem with this potential could be solved to be plane waves, and one can get the information of the conductivity from the solution  following \cite{Kiritsis:2015oxa, Yan2018}. 
We have a hard gap in the conductivity in $M$.\footnote{For choice $Z\sim u^\alpha$ near IR, the leading term $V_\text{eff}\sim \frac{1}{4}\alpha(\alpha+n-2)u^{n-2}$ for $n>2$. In this case, the gapped spectrum exists only when $n\geq 2$. } Using the same argument around \eqref{eq:exax2}, we conclude that the conductivity on $P$ is also always zero.

\subsection{The AdS soliton geometry in Sec. \ref{ss:solcon}}

By taking $Z=\frac{ \sqrt{f} }{u}$ with $f=1-\frac{u^4}{u_0^4}$, the effective potential reads
\bea
V_\text{eff}=\frac{3f}{4u^2}-\frac{f'}{2u}-\frac{f'^2}{16f}+\frac{f''}{4}=\frac{3(u_0^4-u^4)^2-4 u^4u_0^4}{4u_0^4u^2\,(u_0^4-u^4 ) } \,, 
\eea
where the prime is denoting a derivative with respect to $u$. 
From \eqref{eq:newcoor}, the new radial coordinates $\xi$ equals
\bea
\xi=u\, {}_2F_1\left[\frac{1}{4},\frac{1}{2},\frac{5}{4}, \frac{u^4}{u_0^4} \right]
\eea
with $0<\xi<\xi_0=\frac{ \Gamma[\frac{5}{4}] }{ \Gamma[ \frac{3}{4}] } \sqrt{\pi}\, u_0$. Around the tip of the AdS soliton background, i.e., $u\to u_0$, we have
\bea
V_\text{eff}&&= -\frac{1}{4u_0(u_0-u)}+\frac{1}{8u_0^2}+\mathcal{O} (u_0-u) 
\,\\ 
&&=-\frac{1}{4(\xi_0-\xi)^2}+\frac{1}{8u_0^2}+\mathcal{O} (\xi_0-\xi)\,.
\eea
When $u\to 0$, we have $V_\text{eff}=\frac{3}{4u^2}+\mathcal{O}(u^2)$. For this type of effective potential, the solution near the tip behaves as 
\be
v \sim c_1 (\xi_0-\xi)^{1/2}\,,~~~~{\rm when ~} \xi\to \xi_0 
\ee
which is regular and real. This is related to the fact that there is no horizon at $u\to u_0$. In this way, we have real solutions from this Schr\"{o}dinger problem. From \eqref{eq:appflu}, the solution of $a_y$ at $\omega\to 0$ is normalizable, and, therefore, $\omega\to 0$ is not a pole of the system. We have the conductivity with a sum of discrete poles.

\section{Derivation of \eqref{eq:effent}}
\label{app:ree}

In this appendix, we show the detailed derivation of \eqref{eq:effent} following \cite{Myers:2012ed}. The holographic entanglement entropy is given by \eqref{eq:ee1}, and we have a conserved charge as shown in \eqref{eq:consq}, i.e., 
$u^2 \sqrt{1+\frac{u'^2}{f(u)}}=C^{-1}$
with $C^{-1}=u_t^2=u_*^2\sqrt{1+c^2}$ and $u'=\partial u/\partial x$.

Note that the cutoff near the boundary, the profile of the minimal surface, and the coordinates of intersecting point between the extremal surface and the EOW brane $Q$ are all functions of $\ell$, i.e., $\epsilon=\epsilon(\ell)$, $u=u(x,\ell)$, and $x_*=x_*(\ell)$. Performing the variation of  \eqref{eq:ee1} with respect to $\ell$, we can obtain
\begin{align}
\begin{split}\label{eq:varS1}
    \frac{\partial S}{\partial \ell} = \frac{C L}{2G}\left[ \left(1+\frac{u'^2}{f}\right)\bigg|_{x=\ell-\epsilon} \left(1-\frac{d\epsilon}{d\ell}\right) - (1+c^2) \frac{dx_*}{d\ell}
    +\left( \frac{u'}{f} \frac{\partial u}{\partial\ell}\right)\bigg|^{x=\ell-\epsilon}_{x=x_*}\right] \,,
\end{split}
\end{align}
where we have used the equation of motion for the extremal surface:  
\begin{align}
    u''-\left(\frac{f'}{2f}-\frac{2}{u}\right) u'^2+ \frac{2f}{u}=0\,.
\end{align}
Using the fact that the UV cutoff $u_c=u(\ell-\epsilon,\ell)$ is fixed, we have
\begin{align}\label{eq:vaell}
    \left[ u'\left(1-\frac{d\epsilon}{d\ell}\right)+\frac{\partial u}{\partial \ell} \right]\bigg |_{x=\ell-\epsilon}=0\,.
\end{align}

Substituting \eqref{eq:vaell} into \eqref{eq:varS1} we obtain
\begin{align}\label{eq:varS2}
    \frac{\partial S}{\partial \ell} = \frac{C L}{2G}\left[ -\frac{1}{u'}\frac{\partial u}{\partial \ell}\bigg|_{x=\ell-\epsilon} - (1+c^2) \frac{dx_*}{d\ell} +\frac{c}{\sqrt{f}}\frac{\partial u}{\partial \ell}\bigg|_{x=x_*} \right]\,,
\end{align}
where the last term is from $u'(x_*)=-c\sqrt{f(u_*)}$. The first term in the brackets can be calculated from the information near the boundary. From \eqref{eq:uto01}, we know that, when $u\to 0$, 
\begin{align}
    du = - \frac{u_t^2}{u^2} dx + \frac{u_t^2}{u^2} d\ell\,.
\end{align}
 Therefore, we have 
\begin{align}\label{eq:1}
    \frac{1}{u'}\frac{\partial u}{\partial \ell}\bigg|_{x=\ell-\epsilon}=-1\,.
\end{align}

We can calculate the third term in the brackets in \eqref{eq:varS2} by using the same trick as \eqref{eq:vaell}. Using the relation $u(x_*,\ell)=u_*$ and keeping in mind that $ du_*/dx_*=\sqrt{f(u_*)}/c$, we have
\begin{align}\label{eq:varxs}
    \frac{\partial u}{\partial \ell}\bigg|_{x=x_*} = \frac{du_*}{d\ell}-u'(x_*)\frac{dx_*}{d\ell} = \frac{\sqrt{f(u_*)}}{c}(1+c^2) \frac{dx_*}{d\ell}\,.
\end{align}
Using this expression and \eqref{eq:1}, we obtain
\begin{align}
     \frac{\partial S}{\partial \ell} = \frac{C L}{2G} = \frac{ L}{2G} \frac{1}{u_t^2}\,.
\end{align}

Then the renormalized entanglement
entropy can be expressed the same as \eqref{eq:effent}, i.e., 
\begin{align}
    \mathcal F = \frac{\ell^2}{2L} \frac{\partial S}{\partial \ell} = \frac{ 1}{4G} \frac{\ell^2}{u_t^2}\,.
\end{align}
The expressions \eqref{eq:effent2} and \eqref{eq:effentsol} can be obtained by following the same procedure.

\newpage

\end{document}